\documentclass{aa}
\usepackage{natbib,twoopt}
\usepackage{graphicx}
\usepackage[varg]{txfonts}
\usepackage[usenames,dvipsnames,svgnames,table]{xcolor}

\newcommand{\SLIM}{{\sf SLIM}}

\newcommand{\crdb}{{\sc crdb}}
\newcommand{\dragon}{{\sc dragon}}
\newcommand{\galprop}{{\sc galprop}}

\newcommand{\usine}{{\sc usine}}

\newcommand{\xsGalxii}{{\tt Galp-opt12}}
\newcommand{\xsGalxxii}{{\tt Galp-opt22}}
\newcommand{\optxii}{{\tt OPT12}}
\newcommand{\optxiiupxxii}{{\tt OPT12up22}}
\newcommand{\optxxii}{{\tt OPT22}}

\bibpunct{(}{)}{;}{a}{}{,} % to follow the A&A style...

\definecolor{light-gray}{gray}{0.95}
\definecolor{dark-gray}{gray}{0.4}

\usepackage[
            colorlinks=true,%
            citecolor=blue,
            pdfauthor={Maurin et al.},%
            pdftitle={A simple determination of the halo size from 10Be/9Be data}%
           ]{hyperref}
\begin{document}

\input epsf
\title{A simple determination of the halo size from $\mathrm{^{10}Be/^9Be}$ data}
%\subtitle{}
%\titlerunning{}

\author{
  D. Maurin\inst{1}\thanks{\url{david.maurin@lpsc.in2p3.fr}}
  \and E. Ferronato Bueno\inst{2}%\thanks{\url{e.ferronato.bueno@rug.nl}}
  \and L. Derome\inst{1}%\thanks{\url{laurent.derome@lpsc.in2p3.fr}}
}

\authorrunning{Maurin et al.}

\institute{
LPSC, Universit\'e Grenoble Alpes, CNRS/IN2P3, 53 avenue des Martyrs, 38026 Grenoble, France
\and Kapteyn Astronomical Institute, University of Groningen, Landleven 12, 9747 AD Groningen, The Netherlands
}

\date{Received / Accepted}% It is always \today, today,
%  but any date may be explicitly specified

\abstract
% {Text of context}
{The AMS-02 and HELIX experiments should soon provide $\mathrm{^{10}Be/^9Be}$ cosmic-ray data of unprecedented precision.}
% {Text of aims}
{We propose an analytical formula to quickly and accurately determine $L$ from these data.
}
% {Text of methods}
{Our formula is validated against the full calculation performed with the propagation code \usine{}. We compare the constraints on $L$ set by Be/B and $\mathrm{^{10}Be/^9Be}$, relying on updated sets of production cross-sections.}
% {Text of results}
{The best-fit $L$ from AMS-02 Be/B data is shifted from 5~kpc to 3.8~kpc when using the updated cross-sections. We obtained consistent results from the Be/B analysis with \usine{}, $L=3.8^{+2.8}_{-1.6}$~kpc (data and cross-section uncertainties), and from the analysis of $\mathrm{^{10}Be/^9Be}$ data with the simplified formula, $L=4.7\pm0.6$~(data uncertainties)~$\pm2$~(cross-section uncertainties)~kpc. The analytical formula indicates that improvements on $L$  thanks to future data will be limited by production cross-section uncertainties, unless either$\mathrm{^{10}Be/^9Be}$ measurements are extended up to several tens of GeV/n, or nuclear data for the production of $^{10}$Be and $^{9}$Be are improved; new data for the production cross-section of $^{16}$O into Be isotopes above a few GeV/n are especially desired.}
% [Text of perspective} -> Optional
{}

\maketitle
%\setcounter{tocdepth}{2}
%\tableofcontents

%_____________________________________________________________________________
%_____________________________________________________________________________
\section{Introduction}

In the 1950s, \citet{1958PThPS...6....1H} realised that the $^{10}$Be radioactive secondary isotope could be used as a clock to determine the cosmic-ray (CR) age \citep[e.g.][]{1990PhR...191..351S}. In modern CR propagation models, the radioactive clocks are used to determine the halo size $L$ of the Galaxy \citep[e.g.][]{2002A&A...381..539D}. Besides the motivation for a better characterisation of the transport parameters in the Galaxy, the determination of $L$ is also crucial for setting constraints on dark matter from indirect detection of anti-particles \citep{Donato2004,2021PhRvD.104h3005G}.

Many studies have focused on the most abundant and lightest CR clock, $^{10}$Be \citep{1988SSRv...46..205S}, either via the isotopic $\mathrm{^{10}Be/^9Be}$ ratio \citep[e.g.][]{2001ICRC....5.1836M,2002A&A...381..539D} or via the elemental Be/B ratio \citep{1975ICRC....2..526O,1998ApJ...506..335W,2010A&A...516A..66P}. In the former case, the impact of $^{10}$Be decay is maximal in the ratio and, moreover, both the isotopes have roughly the same progenitors and are similarly produced in nuclear reactions. In the latter case, the impact of $^{10}$Be decay is lessened in the numerator by the presence of the more abundantly produced stable isotopes ($^7$Be and $^9$Be), even though the presence of the daughter isotope $^{10}$B in the denominator maximises the impact of decay on the ratio; the modelling of the Be/B ratio also involves a larger network or production cross-sections, potentially leading to larger nuclear uncertainties than for the $\mathrm{^{10}Be/^9Be}$ calculation. On the experimental side, isotopic separation up to high energies remains very challenging, and thus using the Be/B ratio to set constraints on $L$ is complementary to using the $\mathrm{^{10}Be/^9Be}$ ratio.

With the advent of the Alpha Magnetic Spectrometer (AMS-02) experiment on board the International Space Station, data of unprecedented precision have been published for Be/B \citep{2018PhRvL.120b1101A}. These data have been recently analysed by \citet{2020A&A...639A..74W}, \citet{2020PhRvD.101b3013E}, and \citet{2021JCAP...03..099D}, who all highlighted the impact of cross-section uncertainties on the determination of $L$, with $\Delta L\sim\pm2.5$~kpc. The situation for the $\mathrm{^{10}Be/^9Be}$ data is also expected to improve very soon, thanks to the AMS-02 and High Energy Light Isotope eXperiment (HELIX) balloon-borne experiments. In this context, we wish to revisit the uncertainties on $L$ originating from nuclear uncertainties, taking advantage of a recent update on the production cross-sections of Be and B isotopes \citep{2022arXiv220300522M}. We also wish to assess the compatibility of the constraints on $L$ brought by Be/B and $\mathrm{^{10}Be/^9Be}$ data. In particular, for the latter, nuclear uncertainties may be an issue that will prevent us from fully benefiting from the forthcoming measurements.

The Be isotopes are of secondary origin, that is they are thought to solely originate from the fragmentation of heavier CR nuclei. This makes the calculation of CR flux ratios of these isotopes dependent on: (i) the known half-life of the CR clock ($t_{1/2}=1.387$~Myr); (ii) the grammage crossed by CRs during their journey through the Galaxy; (iii) the production cross-sections; (iv) the CR fluxes of their progenitors; and finally, (v) the halo size $L$ of the Galaxy. The grammage can be determined from stable secondary-to-primary ratios \citep[e.g.][]{2001ApJ...555..585M}, while the production cross-sections are set from available nuclear parametrisations and data \citep{2018PhRvC..98c4611G,2022arXiv220300522M}, and measured elemental fluxes are available from a wide range of energies from interstellar (IS) \citep{2016ApJ...831...18C} and top-of-atmosphere (TOA) data \citep{1990A&A...233...96E,2013ApJ...770..117L,2021PhR...894....1A,2021PhRvL.126d1104A,2021PhRvL.126h1102A,2021PhRvL.127b1101A}.
We show that the cross-sections and progenitor fluxes can be combined into a single number (denoted ${\cal F}_{\rm HE}$) that can be used as an input to determine $L$ directly from the $\mathrm{^{10}Be/^9Be}$ data. This simplification should prove useful for experimentalists willing to constrain $L$ from their data without the need for an underlying propagation model.

This paper is organised as follows. In Sect.~\ref{sec:ana1D}, we derive a simple analytical formula to calculate the $\mathrm{^{10}Be/^9Be}$ ratio and we validate this formula against the full calculation performed with the propagation code \usine{}; we also introduce the ${\cal F}$ term and its high-energy limit ${\cal F}_{\rm HE}$, which only depends on the production cross-sections and measured elemental fluxes. In Sect.~\ref{sec:halosize}, we compare the limits on $L$ obtained from the analysis of AMS-02 Be/B data or from available $\mathrm{^{10}Be/^9Be}$ data; we highlight how the uncertainties on nuclear production cross-sections translate into an uncertainty on ${\cal F}_{\rm HE}$, which dominates the uncertainty on $L$. We then conclude in Sect.~\ref{sec:conclusions}.

%_____________________________________________________________________________
%_____________________________________________________________________________

\section{Analytical formula for $\mathrm{^{10}Be/^9Be}$}
\label{sec:ana1D}

The framework we consider in this study is the 1D thin disc/thick halo propagation model, which is known to capture all the salient processes of CR transport while remaining simple \citep[e.g.][]{2001ApJ...547..264J}.
This framework has been used and shown to reproduce AMS-02 high-precision data in several recent studies \citep{2019A&A...627A.158D,2019PhRvD..99l3028G,2019PhRvD..99j3023E,2020PhRvD.101b3013E,2020PhRvR...2b3022B,2020A&A...639A..74W,2020A&A...639A.131W,2021PhRvD.103l3010S}. The semi-analytical solutions in the 1D model model are implemented in the \usine{} code \citep{2020CoPhC.24706942M}, which is used here to validate the analytical formula that we propose.

We consider neither convection nor re-acceleration below, as they are not mandatory in order to give an excellent match to AMS-02 Li/C, Be/B, and B/C data \citep{2020A&A...639A.131W,2022arXiv220300522M}.
Further omitting energy redistribution (re-acceleration and continuous losses), the diffusion equation for the differential density $N^j$ of a species $j$ reduces to
\begin{equation}
\left(-K\frac{d^2}{dz^2} + \frac{1}{\gamma\tau^j_0} + 2hn_{\rm ISM}v\sigma^{j+{\rm ISM}}_{\rm inel}\delta(z)\right)N^j(z) = 2h\delta(z){\cal S}^j(E)\,.
\label{eq:1Ddiff}
\end{equation}
The various terms appearing in this equation are: the diffusion coefficient $K(R)$; the lifetime for an unstable species $\tau^j_0$ ($\tau_0\to\infty$ for a stable species); the disc half-width $h$ pinched in a thin slab $\delta(z)$ where the gas and the sources lie; the destruction rate $n_{\rm ISM}v\sigma^{j+{\rm ISM}}_{\rm inel}$ of species $j$ on the interstellar medium (ISM) of density $n_{\rm ISM}$ (with $\sigma_{\rm inel}$ the inelastic cross-section); and a source term ${\cal S}^j(E)$ containing a primary component $Q^{\rm prim}(E)$ for species accelerated at source, and a secondary component $Q^{\rm sec}(E)=\sum_p n_{\rm ISM} v\sigma^{p+{\rm ISM}\to j}_{\rm prod}N^p$ from the nuclear production of $j$ from all progenitors $p$ heavier than $j$.

\subsection{Neglecting energy losses}
\label{sec:Eloss}
\begin{figure}[!t]
   \centering
   \includegraphics[width=\columnwidth]{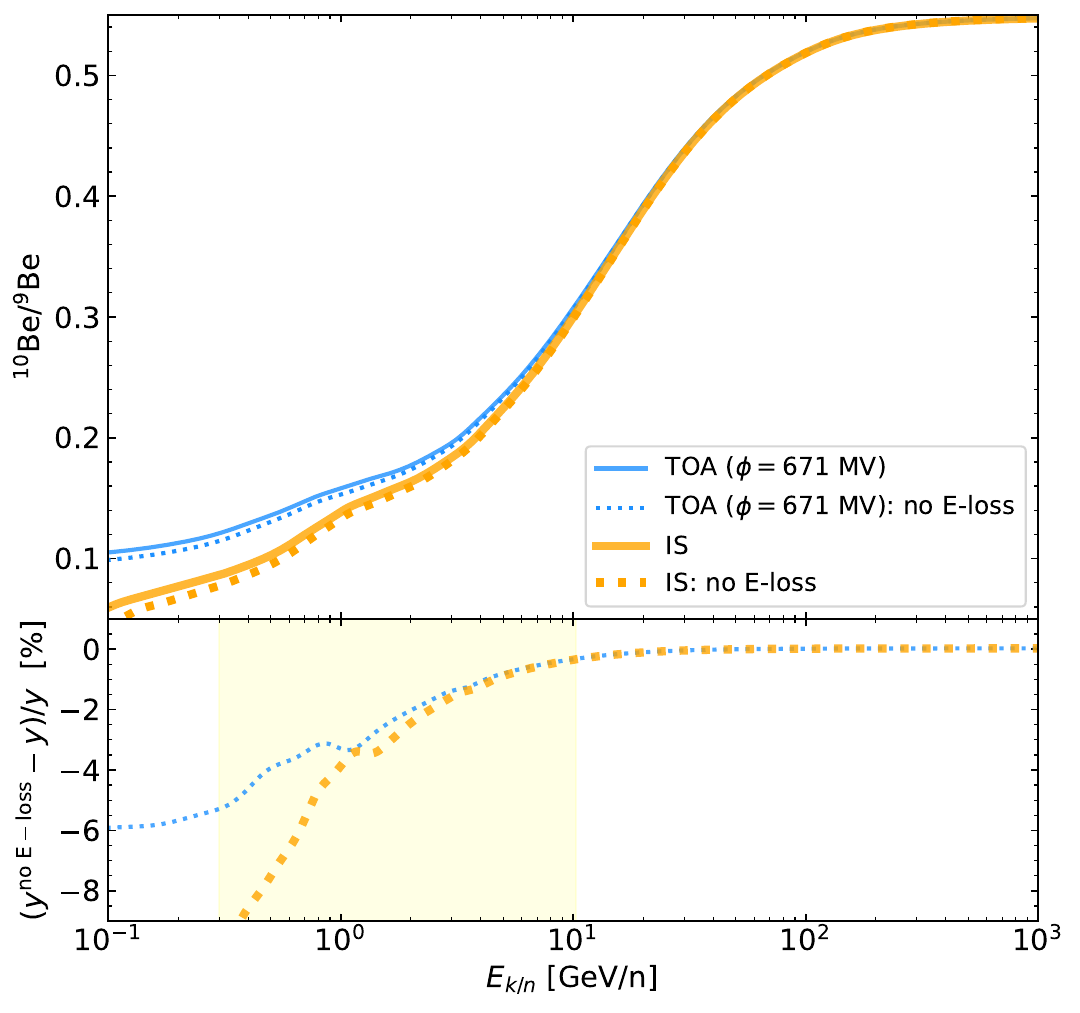}
   \caption{Impact of neglecting energy losses in the $\mathrm{^{10}Be/^9Be}$ ratio as a function of the kinetic energy per nucleon (all curves are from \usine{} runs). {\bf Top panel:} Calculation with energy losses switched on (solid lines) or off (dotted lines) for TOA (blue) or IS (orange) calculations. {\bf Bottom panel:} Relative difference between the `no E loss' and `with E losses' calculations, for the TOA (dotted blue line) and IS (dashed orange line) cases. The range highlighted in yellow corresponds to the region where the AMS-02 and HELIX experiments should measure $\mathrm{^{10}Be/^9Be}$.
   \label{fig:1D_noElosses}}
\end{figure}
We can provide an analytical formula only if we can neglect energy losses. Fig.~\ref{fig:1D_noElosses} shows a comparison of calculations performed with \usine,{} with and without energy losses, based on the best-fit diffusion coefficient and $L$ obtained in  \citet{2020A&A...639A..74W,2020A&A...639A.131W}, for the propagation configuration denoted \SLIM{}.
The comparison is shown for IS quantities (blue lines), but also for TOA quantities (orange lines). The difference between the IS and TOA curves indicates that solar modulation \citep[e.g.][]{2013LRSP...10....3P} is an important ingredient (see Sect.~\ref{sec:solmod}). However, energy losses (a comparison can be made between solid and dotted lines) are not so important. Above a few tens of GeV/n, neglecting energy losses biases the result by a few percent at most for the TOA $\mathrm{^{10}Be/^9Be}$ ratio: this is to be compared to the $15-20\%$ uncertainties expected in forthcoming AMS-02 \citep{2019PhRvL.123r1102A} and HELIX \citep{2019ICRC...36..121P} data. This bias only slightly increases as the energy decreases, meaning that this approximation is equally applicable for data going down to tens of MeV/n (see below). We stress, however, that the $\sim 6\%$ bias gets dangerously close to the best precision current data reach at these energies, with an $\sim 11\%$ precision for ACE-CRIS data \citep{2001ApJ...563..768Y}.

This first comparison shows that, to a first approximation, energy losses can be neglected for the calculation of the TOA $\mathrm{^{10}Be/^9Be}$ ratio above a few tens of MeV/n. The next step, following \citet{1975Ap&SS..32..265P} and \citet{1980Ap&SS..68..295G}, is to calculate an analytical formula for this ratio in the 1D model.

\subsection{Solutions for stable and radioactive species (and ratio)}
\label{sec:anaform}

We are interested here in the $^9$Be (stable) and $^{10}$Be (unstable) isotopes in the disc (i.e. at $z=0$). It is straightforward to solve Eq.~(\ref{eq:1Ddiff}), and we get (we omit all indices for simplicity):
\begin{eqnarray}
N^{\rm stable}(z=0)\!\!\! &=&\frac{{\cal S}(E)}{K\cdot(Lh)^{-1} + n_{\rm ISM}v\sigma_{\rm inel}}\,,\\
N^{\rm rad}(z=0) &=&\frac{{\cal S}(E)}{K\cdot\left(l_{\rm rad}h \cdot \tanh\left(\frac{L}{l_{\rm rad}}\right)\right)^{-1} + n_{\rm ISM}v\sigma_{\rm inel}}\,,
\end{eqnarray}
where we have defined $l_{\rm rad} \equiv \sqrt{\gamma\tau_0 K}$.

To form the ratio of a radioactive to a stable secondary species, it is useful to introduce the diffusion and destruction timescales in the disc, as well as the decay timescale:
\begin{eqnarray}
t_{\rm diff} &=& \frac{Lh}{K}\,, 
\label{eq:t_diff}
\\
t_{\rm inel} &=& \frac{1}{nv\sigma_{\rm inel}}\,,
\label{eq:t_inel}
\\
t_{\rm rad}  &=& \gamma \tau_0 = \gamma \frac{t_{1/2}}{\ln(2)}\,.
\label{eq:t_rad}
\end{eqnarray}
 Using the superscript 10 and 9 to identify quantities calculated for $^{10}$Be and $^9$Be respectively ($t^{10}_{\rm diff}$ and $t^9_{\rm diff}$ are different when calculated at the same $E_{k/n}$ because the diffusion coefficient depends on the rigidity $R=(p/Ze)$),
we then get\begin{equation}
\frac{N^{^{10}}}{N^{^9}} (E_{k/n}) = 
{\cal F} \cdot \frac{t_{\rm diff}^{^{10}}}{t_{\rm diff}^{^9}} \cdot \frac{\displaystyle 1+ \frac{t_{\rm diff}^{^9}}{t_{\rm inel}^{^9}} }{\displaystyle\\\sqrt{\frac{L}{h}\cdot\frac{t_{\rm diff}^{10}}{t_{\rm rad}^{10}}}\cdot\coth\left[\sqrt{\frac{L}{h} \cdot\frac{t_{\rm diff}^{{10}}}{t_{\rm rad}^{{10}}}}\,\right]  + \frac{t_{\rm diff}^{{10}}}{t_{\rm inel}^{^{10}}} }\,,
\label{eq:1Dratio}
\end{equation}
with
\begin{equation}
{\cal F}(E_{k/n}) \equiv \frac{\displaystyle\sum_{p\,\in\,\rm proj}\sum_{t\,\in\,\rm ISM} n_{\rm ISM}^t\,v\,\sigma^{p +t\to ^{10}{\rm Be}}(E_{k/n})\times N^p(E_{k/n})}{\displaystyle\sum_{p\,\in\,\rm proj}\sum_{t\,\in\,\rm ISM} n_{\rm ISM}^t\,v\,\sigma^{p +t\to ^{9}{\rm Be}}(E_{k/n})\times N^p(E_{k/n})}\,.
\label{eq:def_F}
\end{equation}
In this last expression, we put back the summation indices to highlight the fact that ${\cal F}$ only depends on the fragmentation cross-sections into the Be isotopes and the IS progenitor fluxes $\psi^{\rm IS}$; we recall that $\psi^{\rm IS}=Nv/(4\pi)$, because CR fluxes are quasi-isotropic.

For all practical calculations in what follows, the index $t$ runs over H and He ($90\%$ and $10\%$ in number, respectively), whereas the index $p$ runs over all CR species up to $^{56}$Fe. We take $n_{\rm ISM}=1$~g~cm$^{-3}$, but we note that the gas density disappears in ${\cal F}(E_{k/n})$. We come back to Eq.~(\ref{eq:def_F}) and discuss it in detail in Sect.~\ref{sec:Fcst}.

\subsection{Accounting for solar modulation}
\label{sec:solmod}
In practice, the data correspond to TOA quantities, whereas the above formulae correspond to IS quantities. Also, as seen in Fig.~\ref{fig:1D_noElosses} (a comparison can be made between the orange and blue lines in the top panel), the solar modulation effect cannot be neglected. Although we cannot directly modulate a ratio, after a bit of tweaking, we can obtain a simple and accurate enough formula.

In the force-field approximation used in our calculations, TOA and IS quantities are related by  \citep{1967ApJ...149L.115G,1968ApJ...154.1011G}
\begin{eqnarray}
  \psi^{\rm TOA}\left(E^{\rm TOA}\right) &=& \left(\frac{p^{\rm TOA}}{p^{\rm IS}}\right)^2 \psi^{\rm IS}\left(E^{\rm IS}\right)\,,\\
  E_{k/n}^{\rm TOA}&=&E_{k/n}^{\rm IS}-\frac{Z}{A}\phi\,,
\end{eqnarray}
where $\phi$ is the solar modulation level.
For our purpose, we need to calculate the flux ratio at the same TOA kinetic energy per nucleon $E_{k/n}^{\rm TOA}$, whereas solar modulation connects fluxes at total energy. Writing
\begin{eqnarray}
  E^{\rm IS}_{k/n,\,X} &=&  E^{\rm TOA}_{k/n} + \frac{Z_X}{A_X}\phi\,,\nonumber\\
  \left(p^{TOA}_X\right)^2 &=& E^{\rm TOA}_{k/n}A_X\left(E^{\rm TOA}_{k/n}A_X+2m_X\right)\,,\nonumber\\
  \left(p^{IS}_X\right)^2 &=& \left(E^{\rm IS}_{k/n,\,X\,}A_X+m_X\right)^2 -m_X^2\,,\nonumber
\end{eqnarray}
with the subscript $X$ referring to $^{10}$Be or $^9$Be, we get
\begin{equation}
  \frac{\psi_{10}^{\rm TOA}(E_{k/n}^{\rm TOA})}{\psi_9^{\rm TOA}(E_{k/n}^{\rm TOA})}
   \!=\! \left(\frac{\!p_{10}^{\rm TOA}\!}{p_{10}^{\rm IS}}\right)^2
  \!\!\cdot \left(\frac{p_9^{\rm IS}}{\!p_9^{\rm TOA}\!}\right)^2
  \!\cdot \frac{\psi_{10}^{\rm IS}(E^{\rm IS}_{k/n,\,10\,})}{\psi_9^{\rm IS}(E^{\rm IS}_{k/n,\,10\,})}
   \cdot  \frac{\psi_9^{\rm IS}(E^{\rm IS}_{k/n,\,10\,})}{\psi_9^{\rm IS}(E^{\rm IS}_{k/n,\,9\,})}\,.
\end{equation}
In order to form the $^{10}$Be to $^9$Be flux ratio at the same IS kinetic energy per nucleon (next-to-last term in the equation), we need an extra factor (last term) corresponding to the ratio of the $^9$Be flux calculated at two different IS energies. For a good approximation of this flux, we can take advantage of Voyager data taken outside the solar cavity and TOA data at higher energy. In practice, we fit a log-log polynomial on the IS Be flux models shown in Fig.~4 of \citet{2016ApJ...831...18C}, and defining $x\equiv\log_{10}[E_{k/n}/(1~{\rm GeV/n})]$,
we have\begin{equation}
   \log_{10}\left(\psi_9^{\rm IS}\right) = -0.036 -1.283x -0.921x^2 +0.0078x^3 +0.05x^4.
   \label{eq:fluxBe}
\end{equation}
The normalisation of the flux is not important as our calculation involves ratios of $^9$Be fluxes, and we assume that the energy dependence for $^9$Be is the same as that for Be.

Finally, using the above equations and recalling $\psi = N v/(4\pi)$, the formula to calculate the TOA ratio from the IS one is given by
\begin{equation}
  \frac{\psi_{10}^{\rm TOA}(E_{k/n}^{\rm TOA})}{\psi_9^{\rm TOA}(E_{k/n}^{\rm TOA})} = 
   \frac{v_{10}}{v_9}\times \frac{N^{\rm IS}_{10}}{N^{\rm IS}_9} (E_{k/n}^{\rm TOA} + \frac{Z_{10}}{A_{10}}\phi)
   \times {\cal C}_{\rm prefact} 
   \times {\cal C}_{\rm flux}\,,
   \label{eq:TOA2ISforratio}
 \end{equation}
with
\begin{eqnarray}
  {\cal C}_{\rm prefact} & \equiv&  \left(\frac{A_{10}\left(E^{\rm TOA}_{k/n}A_{10}+2m_{10}\right)}{A_9\left(E^{\rm TOA}_{k/n}A_9+2m_9\right)}\right) \nonumber\\
    &\times&  \left(\frac{\left(E^{\rm TOA}_{k/n}A_9+Z_9\phi+m_9\right)^2-m_9^2}{\left(E^{\rm TOA}_{k/n}A_{10}+Z_{10}\phi+m_{10}\right)^2-m_{10}^2}\right)\,,
  \label{eq:corr_prefact} \\
  {\cal C}_{\rm flux} &\equiv& \frac{\psi_9^{\rm IS}\left(E_{k/n}^{\rm TOA} + \frac{Z_9}{A_9}\phi\right)}{\psi_9^{\rm IS}\left(E_{k/n}^{\rm TOA} + \frac{Z_{10}}{A_{10}}\phi\right)},\; {\rm with}~\psi_9^{\rm IS}={\rm Eq.~(12)}\,.
  \label{eq:corr_slope}
 \end{eqnarray}

\begin{figure}[!t]
   \centering
   \includegraphics[width=\columnwidth]{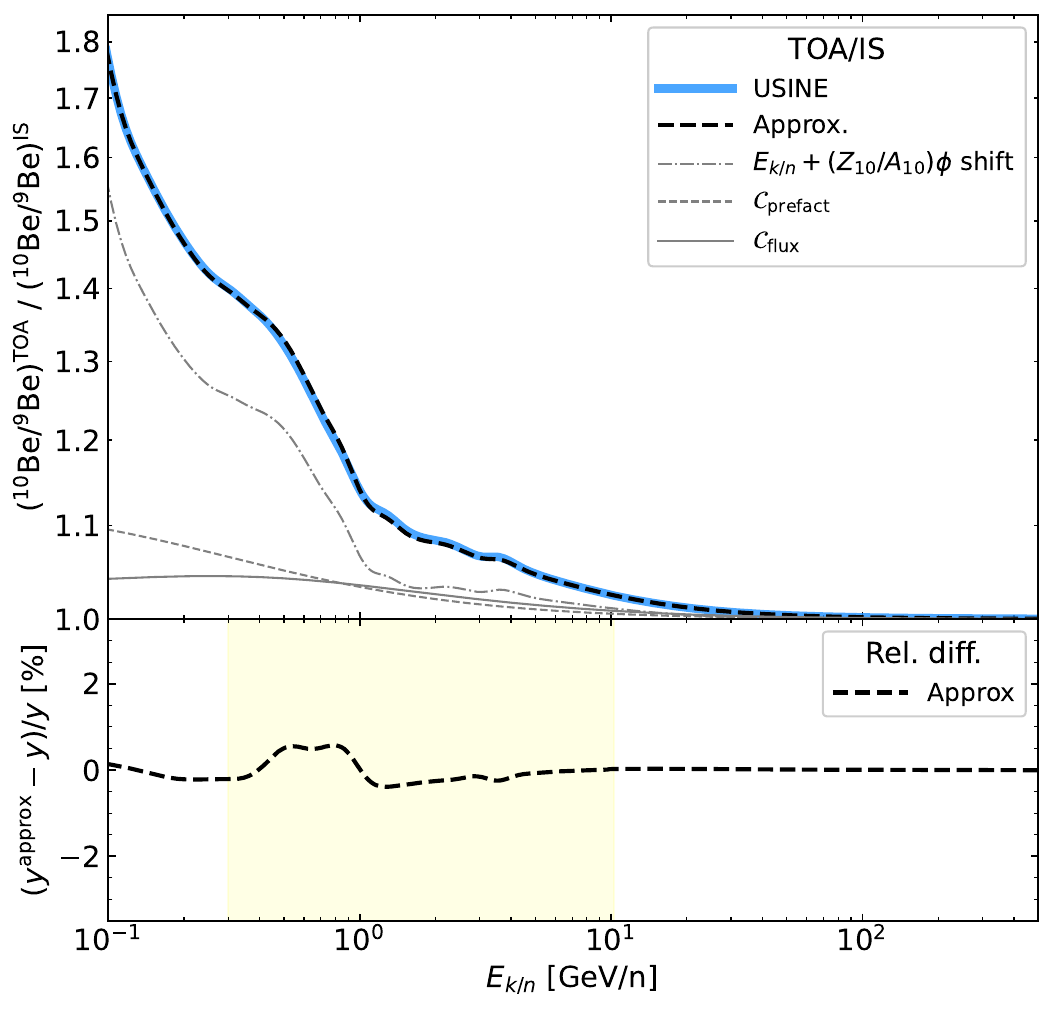}
   \caption{Impact of using an approximate solar modulation, illustrated by the calculation of $(\mathrm{^{10}Be/^9Be})^{\rm TOA}/(\mathrm{^{10}Be/^9Be})^{\rm IS}$ as a function of the kinetic energy per nucleon. {\bf Top panel:} TOA-to-IS ratio for the exact calculation with \usine{} (solid blue line) or using the approximate modulation Eqs.~(\ref{eq:TOA2ISforratio}-\ref{eq:corr_slope}) with Eq.~(\ref{eq:fluxBe}) for the energy dependence of the $^9$Be IS flux (dashed black line). The grey lines show the broken-down corrective terms entering Eq.~(\ref{eq:TOA2ISforratio}). {\bf Bottom panel:} Relative difference between the approximate calculation and the exact calculation. The range highlighted in yellow corresponds to the region where the AMS-02 and HELIX experiments should measure $\mathrm{^{10}Be/^9Be}$.
   \label{fig:1D_solmod}}
\end{figure}
We test the accuracy of the approximate formula against the correct TOA modulation of $\mathrm{^{10}Be/^9Be}$ in the top panel of Fig.~\ref{fig:1D_solmod}. In fact, the figure shows the TOA-to-IS ratio from the exact (solid blue line) and approximate (dashed black line) cases, with the relative difference between the two shown in the bottom panel. The approximation reproduces the exact calculation at the percent level accuracy; the origin of the difference lies in the energy dependence of the IS $^9$Be flux in \usine{} that is not exactly the same as Eq~(\ref{eq:fluxBe}). In the top panel, we also highlight the various corrective terms appearing in the approximate formula, namely ${\cal C}_{\rm prefact}$, ${\cal C}_{\rm flux}$, and `shift'---the TOA ratio must be calculated at IS energy in Eq.~(\ref{eq:TOA2ISforratio}). The dominant correction is from the `shift' term (dash-dotted grey line), with subdominant but still significant corrections from the ${\cal C}_{\rm prefact}$ (dashed grey line) and ${\cal C}_{\rm flux}$ (dotted grey line) terms.

\subsection{${\cal F}$ at high energy and ${\cal F}(E_{k/n})={\cal F}_{\rm HE}$ approximation}
\label{sec:Fcst}

The ${\cal F}$ term defined in Eq.~(\ref{eq:def_F}) is, a priori, an energy-dependent quantity. We recall that it can be directly calculated from the knowledge of measured IS fluxes for $^{10}$Be and $^9$Be progenitors and production cross-sections of these progenitors into the Be isotopes. For IS fluxes, parametrisations for H to Fe elements are available \citep{2019ApJ...887..132S,2020ApJS..250...27B}, based on Voyager data taken at IS energies \citep{2016ApJ...831...18C} and ACE~\citep{2009ApJ...698.1666G,2013ApJ...770..117L}, AMS-02 \citep{2021PhR...894....1A}, and HEAO-3 \citep{1990A&A...233...96E} data taken at TOA energies. For the production cross-sections, several datasets are also publicly available from the \dragon{}\footnote{\url{https://github.com/cosmicrays}} \citep{2018JCAP...07..006E}, \galprop{}\footnote{\url{https://galprop.stanford.edu/}} \citep{2021arXiv211212745P}, and \usine{}\footnote{\url{https://lpsc.in2p3.fr/usine}} \citep{2020CoPhC.24706942M} codes and websites.

Both IS fluxes and production cross-sections are subject to uncertainties that are difficult to quantify. For this reason, we do not provide pre-calculated ${\cal F}$ terms for the various existing (and still evolving) parametrisations. Rather, we show that taking ${\cal F}$ to be a constant provides an even simpler framework for analysing $(\mathrm{^{10}Be/^9Be})^{\rm TOA}$ data, while providing meaningful constraints on $L$.

\paragraph{Asymptotic behaviour at high energy.}

The ${\cal F}$ term is expected to become constant above GeV/n. Indeed, the production cross-sections are generally assumed to be constant above a few GeV/n. Moreover, the main progenitors of $^{10}$Be and $^9$Be at high energy are always primary species with the same energy dependence (species from C to Fe are assumed to share the same source slope), so that this energy dependence cancels out of ${\cal F}$.

An asymptotically constant ${\cal F}={\cal F}_{\rm HE}$ term at high energy (HE) means that the $\mathrm{^{10}Be/^9Be}$ ratio also becomes a constant---$^{10}$Be behaves like a stable species at high energy (boosted decay time $\gamma\tau_0$). This is illustrated in Figure~\ref{fig:1D_noElosses}, where the asymptotic regime is reached above a few hundred GeV/n. Indeed, ${\cal F}_{\rm HE}$ can be directly inferred from model calculations of the high-energy $(\mathrm{^{10}Be/^9Be})_{\rm HE}$ ratio, because Eq~(\ref{eq:1Dratio}) reduces to
\begin{equation}
\left.\frac{^{10}{\rm Be}(E_{\rm k/n})}{\;^{9}{\rm Be}(E_{\rm k/n})}\right|_{\rm HE} \equiv
\left.\frac{\psi_{10}^{\rm TOA}(E_{\rm k/n})}{\psi_9^{\rm TOA}(E_{\rm k/n})}\right|_{\rm HE} \approx \frac{t_{\rm diff}^{10}}{t_{\rm diff}^{9}} \cdot {\cal F}_{\rm HE}  \approx 0.949\, {\cal F}_{\rm HE}.
\label{eq:FtoHE}
\end{equation}
For the last equality, we assumed a high-rigidity diffusion coefficient $K(R)\propto R^\delta$ with $\delta=0.5$ \citep{2019PhRvD..99l3028G,2020A&A...639A.131W,2022arXiv220300522M}, appropriate for a high-energy (HE) regime, taken to be $\gtrsim 100$~GeV/n here\footnote{Asymptotically, $t_{\rm diff}^{10}/t_{\rm diff}^{9} \approx (9/10)^\delta$ because $R$ is evaluated at the same kinetic energy per nucleon for the two isotopes. Our results are insensitive to the exact value used for $\delta$.}.

\paragraph{Assuming ${\cal F}={\cal F}_{\rm HE}$ in the analytical model.}
It is interesting to see whether using ${\cal F}= {\cal F}_{\rm HE}$ in Eq.~(\ref{eq:1Dratio}) provides a good enough approximation to further simplify the fit of $^{10}$Be/$^9$Be data. Departure of ${\cal F}$ from a constant happens for two reasons. First, production cross-sections become energy dependent below a few GeV/n. This effect is mitigated by the fact that data are at TOA energies, and thus correspond to cross-sections evaluated at $E_{k/n}^{\rm IS}\gtrsim E_{k/n}^{\rm TOA} +0.5$~GeV/n. Second, as observed directly in the AMS-02 data \citep{2021PhRvL.126d1104A}, the ratio of primary species becomes energy dependent below $\lesssim 100$~GeV/n. This is because inelastic interactions more strongly impact  heavier species, so that all parents have slightly different energy dependences: the flux ratio of heavier-to-lighter primary elements decreases with decreasing energies \citep{2011A&A...526A.101P,2022arXiv220306479V}. Progenitors of Be isotopes can also be of secondary (e.g. B) or mixed (e.g. N) origin, and thus have different energy dependences compared to purely primary CR fluxes (e.g. O, Si, and Fe). However, to some extent, this variety of energy dependences in the progenitors is mitigated by the fact that the most important parents for $^9$Be and $^{10}$Be are mostly the same species \citep{2018PhRvC..98c4611G}.

\begin{figure}[!t]
   \centering
   \includegraphics[width=\columnwidth]{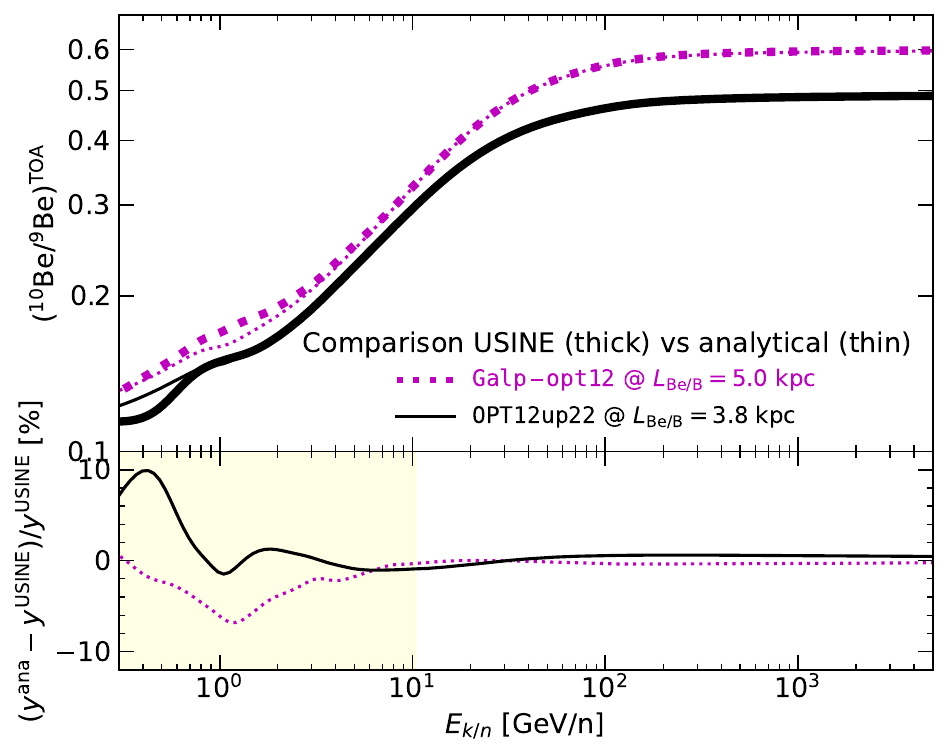}
   \caption{Impact of using the approximation ${\cal F}={\cal F}_{\rm HE}$ for the calculation of $(\mathrm{^{10}Be/^9Be})^{\rm TOA}$ as a function of the kinetic energy per nucleon.
   {\bf Top panel:} Calculation from \usine{} (thin lines) and the analytical model (thick lines) for two different production cross-section sets (associated with different transport parameters and $L$ values, as highlighted in the legend). {\bf Bottom panel:} Relative difference between the approximate calculation and the \usine{} calculation. The range highlighted in yellow corresponds to the region where the AMS-02 and HELIX experiments should measure $\mathrm{^{10}Be/^9Be}$.
   \label{fig:comparison_ratio}}
\end{figure}
In Fig.~\ref{fig:comparison_ratio} we compare the bias incurred by assuming ${\cal F}(E_{k/n})={\cal F}_{\rm HE}$ in the $\mathrm{^{10}Be/^9Be}$ calculation, for two different production cross-section sets: \xsGalxii{} (magenta) and \optxiiupxxii{} (black). These sets are properly introduced and discussed in Sect.~\ref{sec:xs_sets}. The top panel shows the $(\mathrm{^{10}Be/^9Be})^{\rm TOA}$ ratio calculated with \usine{} (thick lines) and with the analytical model assuming ${\cal F}={\cal F}_{\rm HE}$ (thin lines); the bottom panel shows the relative difference between the two\footnote{Both models rely on the same values for the transport parameters and $L$, taken to be the best-fit parameters from the Be/B analysis with \usine{} (see Sect.~\ref{sec:fit_BeB}). In detail, the analytical model uses Eqs~(\ref{eq:TOA2ISforratio}) and~(\ref{eq:1Dratio}), and $t_{\rm diff}$ and $L$ are fixed to the values appropriate for the production cross-section set considered, with ${\cal F}_{\rm HE}$ retrieved from the bottom panel of Fig.~\ref{fig:freeL_BeB} using Eq.~(\ref{eq:FtoHE}), that is ${\cal F}_{\xsGalxii{}}=0.62$ and ${\cal F}_{\optxiiupxxii{}}=0.51$.}.
For both of the cross-section sets, the analytical model (thick lines) recovers the \usine{} calculation (thin lines) at a precision better than $1\%$  above $\gtrsim 4$~GeV/n. Below, a non-trivial energy dependence is seen, dependent on the cross-section set: for \optxiiupxxii{} (solid black line), the agreement is only a few percent, except for a peak difference of $10\%$  at 300~MeV/n (black line in the bottom panel); for \xsGalxii{}, the difference grows slowly, and is above $5\%$ at TOA energies below 100 GeV/n.

\subsection{Consequences for the determination of $L$}

Equations~(\ref{eq:1Dratio}) and (\ref{eq:TOA2ISforratio}), with the full energy dependence accounted for in ${\cal F}$, provide an excellent approximation (few percent precision) for the calculation of $(\mathrm{^{10}Be/^9Be})^{\rm TOA}$ from a few tens of MeV/n up to the highest energies. This approximation is less accurate if we assume ${\cal F}={\cal F}_{\rm HE}$: differences as large as $10\%$ appear, although in a very narrow energy domain around tens or hundreds of MeV/n. This approximation is also more accurate or less accurate, depending on the cross-section set used.

The advantage of the ${\cal F}={\cal F}_{\rm HE}$ approximation is that it allows some ingredients entering the calculation to be further separated and identified.
Indeed, the main terms in Eq.~(\ref{eq:TOA2ISforratio}) are now the timescales in Eqs.~(\ref{eq:t_diff}-\ref{eq:t_rad}) and ${\cal F}_{\rm HE}$. First, $t_{\rm diff}$ corresponds to the confinement time in the disc (or grammage), and it is determined by the analysis of stable secondary-to-primary species \citep[e.g.][]{2020A&A...639A.131W}, that is independently of $L$. Second, we already stressed that ${\cal F}$ can be calculated independently of any propagation model but ${\cal F}_{\rm HE}$ is even simpler, because it is directly related to $(\mathrm{^{10}Be/^9Be})_{\rm HE}$ via Eq.~(\ref{eq:FtoHE}). Third, with $t_{\rm inel}$ calculated from the inelastic cross-sections, this leaves Eq.~(\ref{eq:1Dratio}) exhibiting a single free parameter only, namely the halo size $L$. This means that $L$ can be directly determined from $(\mathrm{^{10}Be/^9Be})^{\rm TOA}$ data, provided ${\cal F}_{\rm HE}$ does not suffer from excessively large uncertainties.  In other words, the determination of $L$ will be limited by the precision at which the factor ${\cal F}_{\rm HE}$ can be determined, which is mostly related to the precision of the production cross-sections of $^9$Be and $^{10}$Be.

%_____________________________________________________________________________
%_____________________________________________________________________________
\section{Halo size determination: Be/B versus $^{10}$Be/$^9$Be}
\label{sec:halosize}

For all the analyses in this paper, whether we run \usine{} or use the analytical formula, we rely on inelastic cross-sections from \citet{1997lrc..reptQ....T,1999STIN...0004259T} and an ISM composed of $90\%$ of H and $10\%$ He (in number). The secondary Be/B or $\mathrm{^{10}Be/^9Be}$ ratios, and in particular the ${\cal F}$ term (or ${\cal F}_{\rm HE}$ discussed above), also crucially depend on the nuclear production cross-section sets considered. As such, we need to discuss them before setting the constraints on $L$.

\subsection{Production cross-section sets}
\label{sec:xs_sets}
In this study we consider four different production cross-section sets, including the \xsGalxii{} set from the \galprop{} team, because it has been the most used set in recent CR analyses \citep{2019PhRvD..99l3028G,2020A&A...639A.131W,2020ApJS..250...27B,2020ApJ...889..167B,2021PhRvD.103j3016K,2021arXiv210803687W}; it is also the set used in our previous effort to determine $L$ from AMS-02 Be/B data \citep{2020A&A...639A..74W}. We also consider the \optxii{}, \optxiiupxxii{}, and \optxxii{} sets, discussed in detail in \citet{2022arXiv220300522M}, with the most-plausible set being \optxiiupxxii{}. These sets correspond to various ways to renormalise the original \galprop{} cross-section options (\xsGalxii{} and \xsGalxxii{}) on the most important production channels \citep{2018PhRvC..98c4611G}, taking advantage of recent nuclear data, and also including the misestimated but important contribution of Fe fragmentation into Li, Be, and B isotopes \citep{2022arXiv220300522M}.

Other production cross-section sets exist, as derived for instance by \citet{2018JCAP...01..055R}, the \dragon{} team \citep{2018JCAP...07..006E}, or from FLUKA \citep{2022JCAP...07..008D}. We do not directly use these other sets in this analysis, but we illustrate in Sect.~\ref{sec:fit_10Be9Be} how their use would impact the determination of $L$.

\subsection{Updated constraints from AMS-02 Be/B data}
\label{sec:fit_BeB}

In this section, the constraints on the halo size $L$ from AMS-02 Be/B data are derived for the four production cross-section sets presented above.
In practice, we repeat the analysis of \citet{2020A&A...639A..74W}, that is we fit AMS-02 Li/C, Be/B, and B/C data, accounting for the covariance matrix of uncertainties on the data, and for nuisance parameters on nuclear cross-section parameters and on the solar modulation level. This allows us to determine the transport parameters along with the halo size $L$---actually $\log_{10}(L/~1{\rm kpc})$ in the fit---, which is the only parameter we discuss here\footnote{ In this paper we do not discuss at all the values of the diffusion parameters (diffusion slope, normalisation, low- and high-rigidity breaks). We recall that they are by-products of the combined analysis of Li/C, Be/C, an B/C data. We stress in particular that each production cross-section set has slightly different transport parameter best-fit values and uncertainties; see \citet{2022arXiv220300522M} for details.}.

\paragraph{$L$ values.}
In Table~\ref{tab:L_BeB} we report the best-fit values (and uncertainties) obtained for $L$ for the various cross-section sets introduced in Sect.~\ref{sec:xs_sets}; for \xsGalxii{} (first line), we directly reproduce the numbers from \citet{2020A&A...639A..74W}. We also report the associated $\chi^2_r$ values, confirming that the cross-section set \optxiiupxxii{} is the one that best fits the data\footnote{This conclusion was reached in \citet{2022arXiv220300522M} for an analysis at fixed $L$, while it is extended here to an analysis at free $L$.}.
\begin{table}[!t]
  \centering
  \caption{Constraints on the halo size $L$ and $1\sigma$ uncertainties from the combined analysis of AMS-02 Li/C, Be/B, and B/C data with \usine{}. We report the reduced $\chi^2$ value for the best-fit (201 data points, 193 degrees of freedom). The different rows show the results for the original \xsGalxii{} cross-section set used in our previous analysis \citep{2020A&A...639A..74W}, and the three updated sets introduced in \citet{2022arXiv220300522M}.}
  \label{tab:L_BeB}
\small
\begin{tabular}{lcc}
  \hline \hline \\[-1em]
  \multicolumn{3}{c}{Fit Li/C+Be/B+B/C with \usine{}} \\
  Cross-section set &       L [kpc]       &$\chi^2_r$ \\
  \hline\\
   \xsGalxii{}      & $5.0^{+3.0}_{-1.8}$ &   1.20  \\[2pt]
   \optxii{}        & $5.6^{+5.6}_{-2.5}$ &   1.16  \\[2pt]
   \optxiiupxxii{}  & $3.8^{+2.8}_{-1.6}$ &   1.13  \\[2pt]
   \optxxii{}       & $4.6^{+4.0}_{-2.1}$ &   1.20  \\[2pt]
  \hline\\[-15pt] 
\end{tabular}
\end{table}

In Table~\ref{tab:L_BeB} we see that the choice of the production cross-section set strongly impacts the best-fit halo size, but less so their uncertainties (which are at least twice as large than the difference between the best-fit values): the best-fit constraint on $L$ moves from $L_\xsGalxii{}=5.0^{+3.0}_{-1.8}$ \citep{2020A&A...639A..74W} to $L_\optxiiupxxii{}=3.8^{+2.8}_{-1.6}$ (this analysis), with a significantly improved $\chi^2$.
These differences arise partly from the fact that the updated cross-section sets (\optxii{}, \optxiiupxxii{}, and \optxxii{}) lead to differently enhanced productions of Li (and to a lesser extent Be) with respect to the original \xsGalxii{}, and thus to a different baseline grammage to reproduce the same secondary-to-primary data (see Fig.~10 of \citealt{2022arXiv220300522M}). However, the main difference is related to the modified cross-section for the production of $^{10}$Be: looking at the columns labelled $^9$Be, $^{10}$Be, and $^{10}$B in Table~1 of \citet{2022arXiv220300522M}, the numbers correspond to the rescaling applied to obtain \optxiiupxxii{} from \xsGalxii{} cross-sections, and they vary significantly for the various parents (rows), especially $^{56}$Fe, $^{16}$O, and $^{12}$C. Moreover, Table~2 in \citet{2022arXiv220300522M} illustrates that, depending on the rescaling procedure used and scatter on nuclear data, the production cross-section for $^{10}$Be can vary by a factor of $\gtrsim 50\%$ for some reactions (as illustrated in the right panel of their Fig.~6).

\paragraph{Be/B best-fit ratio and $1\sigma$ contour.}

\begin{figure}[!t]
   \centering
   \includegraphics[width=\columnwidth]{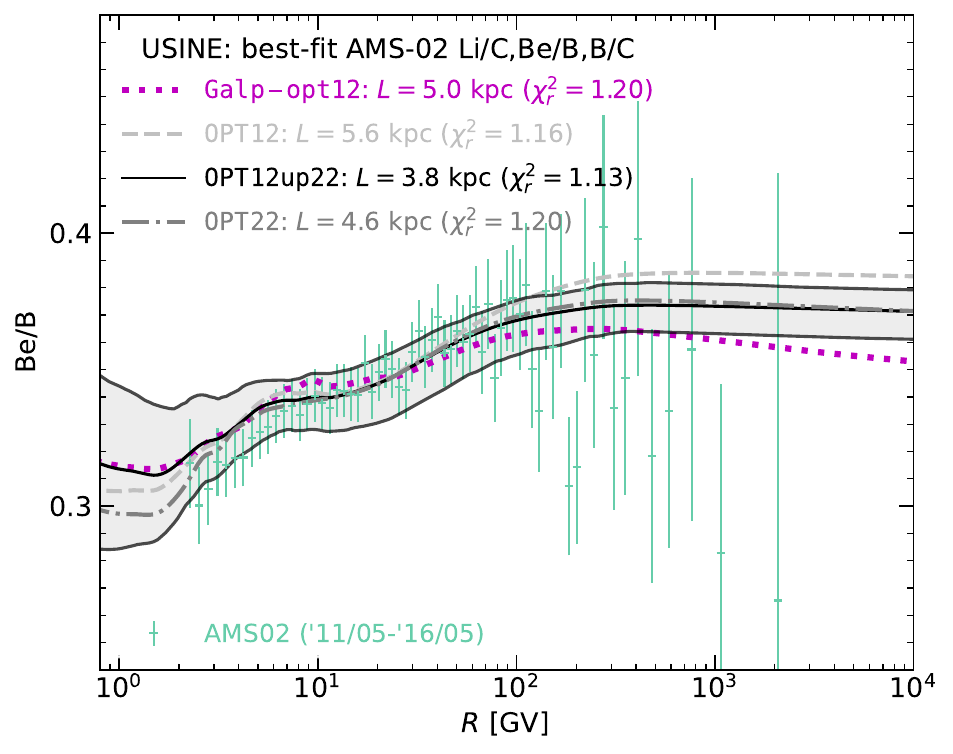}
   \includegraphics[width=\columnwidth]{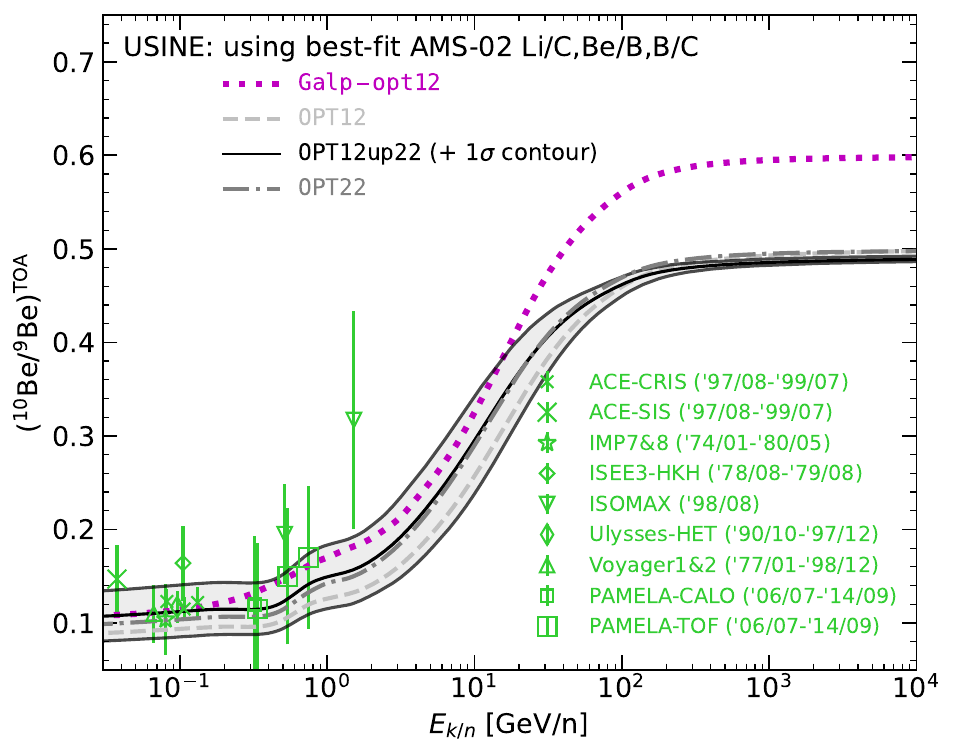}
   \caption{Best-fit models from the combined analysis of AMS-02 Li/C, Be/B, and B/C data with \usine{} (see left-hand side of Table~\ref{tab:L_BeB}). The four line styles correspond to the four production cross-section sets discussed in Sect.~\ref{sec:xs_sets} (\xsGalxii{}, \optxii{}, \optxiiupxxii{}, and \optxxii{}). We also show $1\sigma$ CLs, but for the \optxiiupxxii{} set only (solid black lines), the other sets giving similar contours.
   {\bf Top panel:} Be/B as a function of rigidity, along with the AMS-02 data \citep{2018PhRvL.120b1101A} used for the fit.
   {\bf Bottom panel:} $\mathrm{^{10}Be/^9Be}$ as a function of kinetic energy per nucleon for the same transport parameters and $L$ as above (i.e. from best-fit to Li/C, Be/B, and B/C data only, no fit to $\mathrm{^{10}Be/^9Be}$). A compilation of data (green symbols) is shown for illustrative purpose.
   \label{fig:freeL_BeB}}
\end{figure}
It is interesting to inspect the best-fit values and contours obtained for the Be/B ratio. This is shown in the top panel of Fig.~\ref{fig:freeL_BeB}, along with the AMS-02 data used for the fit. The difference between the use of the original cross-section set \xsGalxii{} (dotted magenta line) and the updated ones (grey and dark lines) is very mild, except above a few hundred GeV/n. To understand the origin of this $\lesssim5\%$ difference, we again have to refer to the underlying cross-section reactions \citep{2022arXiv220300522M}: in the updated cross-section sets (\optxii{}, \optxiiupxxii{}, and \optxxii{}), the production of Be, and in particular $^7$Be (which is the dominant isotope, see Fig.~13 in \citealt{2022arXiv220300522M}), is increased by the presence of Fe fragmentation compared to the less important extra production of B above $\sim 10~$GV---the importance of Fe in the production of Be and B is shown in their Fig.~8. It is nice, however, to see that the overall shape of the ratio, and the small features seen in the data at low rigidity, seem to be even better reproduced with the updated cross-section sets.

\paragraph{Resulting $\mathrm{^{10}Be/^9Be}$ best-fit ratio and $1\sigma$ contour.}
It is also interesting to look at what these best-fit models predict for the $\mathrm{^{10}Be/^9Be}$ ratio. This is shown in the bottom panel of Fig.~\ref{fig:freeL_BeB}, where we see that all models are in fair agreement with the existing data (not used in the fit in this section). However, there are significant differences between the models, and the origin of these differences is not the same at low and high energy.

At high energy, the $\mathrm{^{10}Be/^9Be}$ ratio becomes constant. As already stressed (see Sect.~\ref{sec:ana1D}), this ratio is independent of $L$ and $t_{\rm diff}$, but very sensitive to the production cross-section values of the most-important progenitors via ${\cal F}_{\rm HE}$ (see Sect. \ref{sec:Fcst}). The differences seen between \xsGalxii{} (dotted magenta line) and the updated sets  \optxii{} (dashed grey line), \optxiiupxxii{} (solid black line), and \optxxii{} (dash-dotted grey line), simply reflect the changed proportion of produced $^9$Be and $^{10}$Be---see discussion from the previous paragraph.

At low energy, where decay dominates for $^{10}$Be, we see differences also between the three updated cross-section sets. In this regime, the origin of the differences, as can be read off Eq.~(\ref{eq:1Dratio}), is related to both $t^{10}_{\rm diff}/t^{10}_{\rm rad}$ and $L$. Actually, it must be only related to $t^{10}_{\rm diff}$ and $L$ as all other terms entering the analytical equation remain unchanged from one cross-section set to another. To go into more details, moving from \optxii{} to \optxxii{}, there is an increased overall production of Li, Be, and B \citep{2022arXiv220300522M}, so that a lower grammage is necessary to produce the same amount of measured secondaries. However, this increase is at the $\sim 20\%$ level (see Fig.~10 in \citealt{2022arXiv220300522M}), compared to the much larger scatter observed on the $L$ best-fit values (see Table~\ref{tab:L_BeB}); it is not clear which of the previous two effects dominates the scatter on the $(\mathrm{^{10}Be/^9Be})^{\rm TOA}$ isotopic ratio. This scatter is smaller than the width of the $1\sigma$ envelope (grey-shaded area) calculated from the covariance matrix of the best-fit parameters (dominated by $L$ uncertainties). At growing energies, the contours shrink as the calculation becomes independent of $L$ and of the transport coefficient.

\subsection{Constraints from $^{10}$Be/$^9$Be data with the analytical model}
\label{sec:fit_10Be9Be}

We can now move on to the analysis of the constraints set on $L$ by $(\mathrm{^{10}Be/^9Be})^{\rm TOA}$ data, using the analytical model. Below we compare the results obtained using the full energy dependence in ${\cal F}(E_{k/n})$ or the approximation ${\cal F}(E_{k/n})={\cal F}_{\rm HE}$ (i.e. a constant).

\paragraph{$L$ values.}

To perform the fit on $L$ for a given production set, we calculate ${\cal F}$ (or ${\cal F}_{\rm HE}$) from \usine{} IS fluxes and production cross-sections, and $t_{\rm diff}$ is calculated from the transport parameters given in \citet{2022arXiv220300522M}\footnote{These transport parameters are obtained at $L=5$~kpc, but we recall that the diffusion time $t_{\rm diff}=Lh/K$ is independent of the value $L$ at which it is evaluated, since $K/L$ is the quantity constrained by secondary-to-primary ratios.}; we report the values for ${\cal F}_{\rm HE}$ and $t_{\rm diff}$ in Table~\ref{tab:L_10Be9Be}.
The $\mathrm{^{10}Be/^9Be}$ data used are those extracted from \crdb{}\footnote{\url{https://lpsc.in2p3.fr/crdb}} \citep{2014A&A...569A..32M,2020Univ....6..102M}, namely ACE \citep{2001ApJ...563..768Y}, IMP7\&8 \citep{1981ICRC....2...72G}, ISEE3 \citep{1980ApJ...239L.139W}, ISOMAX \citep{2004ApJ...611..892H}, PAMELA-CALO and TOF \citep{2021Univ....7..183N}, Ulysses \citep{1998ApJ...501L..59C}, and Voyager1\&2 \citep{1999ICRC....3...41L}. For each dataset, we apply a different modulation level $\phi$, also retrieved from \crdb{}. The $\phi$ values are based on the analysis of neutron monitor data \citep{2017AdSpR..60..833G}, from a careful calibration of the H and He fluxes \citep{2016A&A...591A..94G} and neutron monitor response \citep{2015AdSpR..55..363M}. We have $\phi_{\rm ACE} = 609$~MV, $\phi_{\rm IMP7\&8} = 666$~MV, $\phi_{\rm ISEE} = 741$~MV, $\phi_{\rm ISOMAX} = 597$~MV, $\phi_{\rm PAMELA} = 783$~MV, and $\phi_{\rm Ulysses} = 727$~MV; for Voyager1\&2, we keep the value $\phi = 460$~MV quoted in the original reference \citep{1999ICRC....3...41L}, corresponding to an effective modulation level calculated for the spacecraft at an average distance of 30~AU from the Sun.

\begin{figure}[!t]
   \centering
   \includegraphics[width=\columnwidth]{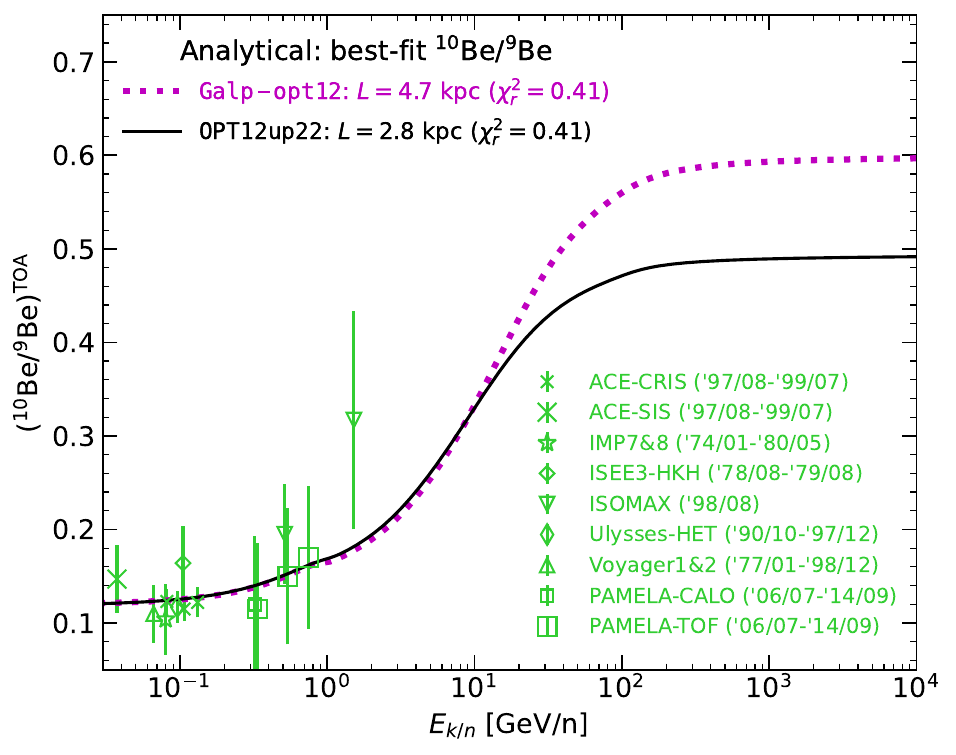}
   \caption{Best-fit model on $\mathrm{^{10}Be/^9Be}$ data using the analytical model, for ${\cal F}$ and $t_{\rm diff}$ input values estimated from different publications: \xsGalxii{} (dotted magenta line) and \optxiiupxxii{} (solid black line) are taken from \citet{2022arXiv220300522M}.
   The models are shown for $\phi=700$~MV, typically corresponding to the average modulation of all the data (green symbols), whose references are given in the text.
   \label{fig:freeL_10Be9Be}}
\end{figure}

\begin{table*}[!t]
  \centering
  \caption{Constraints on the halo size $L$ from the analysis of $\mathrm{^{10}Be/^9Be}$ data with the analytical formula. The first three columns report the production cross-section sets and their associated diffusion time $t_{\rm diff}$, and the asymptotic high-energy value ${\cal F}_{\rm He}$ calculated from \citet{2022arXiv220300522M}. The next columns report the best-fit $L$ (and uncertainties) for the analysis based on the approximation ${\cal F}={\cal F}_{\rm HE}$ or from the use of the full energy dependence in ${\cal F}(E_{k/n})$. We also show, for the two cases, the associated reduced $\chi^2$ (14 data points, 13 degrees of freedom).}
  \label{tab:L_10Be9Be}
\small
\begin{tabular}{lcc@{\hskip1cm}cc@{\hskip1cm}cc}
  \hline \hline \\[-1em]
   \multicolumn{3}{c}{Input configuration}&\multicolumn{4}{c}{\hspace{-0.5cm}Fit $\mathrm{^{10}Be/^9Be}$ (analytical)} \\[3pt]
  Cross-section set &  ${\cal F}_{\rm HE}$ & $t_{\rm diff}\left(R~{\rm [GV]}\right)$ &\!\!\!$L$ using ${\cal F}_{\rm HE}$\!\!\!& $\chi^2_r$ &$\!\!\!\!\!L$ using ${\cal F}(E_{k/n})\!\!\!\!\!$&$\chi^2_r$\\
                    &  [-] & [Myr] & [kpc] &\!\!\!(dof=13)\!\!\! & [kpc] & \!\!\!(dof=13)\!\!\!\\
  \hline\\[-6pt]
  \xsGalxii{}
      & 0.62 & $12.85 \, \beta^{-1} R^{-0.51}\left(1+\left(\frac{R}{4.53}\right)^{-25}\right)^{-0.05}\left(1+\left(\frac{R}{246.7}\right)^{4.54}\right)^{0.041}$
      & $4.7\pm0.6$ & 0.41 & $5.1\pm0.6$ & 0.46\\[12pt]
  \optxiiupxxii{}
      & 0.51 & $10.68 \, \beta^{-1} R^{-0.50}\left(1+\left(\frac{R}{4.67}\right)^{-17.24}\right)^{-0.05}\left(1+\left(\frac{R}{246.7}\right)^{4.54}\right)^{0.041}$
      & $2.8\pm0.4$ & 0.41 & $2.8\pm0.3$ & 0.40\\[12pt]
%  {\tt DRAGON2} [De21]
%      & 0.52 & $5.143\,\beta^{0.6} R^{-0.42}$
%      & $3.4\pm0.4$ & 0.39 &    \dots    &\dots\\[12pt]%L_pub=6.76 kpc
%  {\tt WEBBER} [De21]
%      & 0.33 & $4.860\,\beta^{0.25} R^{-0.42}$
%      & $1.1\pm0.2$ & 0.45 &    \dots    &\dots\\[12pt]%L_pub=2.07 kpc
  \hline\\[-15pt]
\end{tabular}
\end{table*}
We show the best-fit results (and uncertainties) for $L$ in Table~\ref{tab:L_10Be9Be} using either the approximation ${\cal F}={\cal F}_{\rm HE}$ or using the full energy dependence in ${\cal F}(E_{k/n})$. Both approaches lead to consistent $L$ values within the uncertainties.  The latter originate from the $\mathrm{^{10}Be/^9Be}$ data uncertainties, and we have
$$
(\Delta L/L)^{\mathrm{^{10}Be/^9Be}}_{\rm CR~data}\approx5-10\%.
$$
We find that the bias on $L$ from the ${\cal F}={\cal F}_{\rm HE}$ approximation is smaller than $(\Delta L)_{\rm CR~data}$ and dependent on the cross-section set used. In the rest of the section, we focus on results obtained with the approximation (i.e. using ${\cal F}_{\rm HE}$ only). The associated best-fit models (modulated at 700~MV) are shown in Fig.~\ref{fig:freeL_10Be9Be} for illustration. They all go nicely through the data for all production sets considered, reflecting the fact that $\chi^2_r \approx 0.4$ for all these configurations.

\paragraph{Propagation of cross-section uncertainties on $L$.}
As for the Be/B analysis, using the cross-section set \optxiiupxxii{} leads to a smaller best-fit value $L$ than using \xsGalxii{} (a comparison can be made between the numbers in Table~\ref{tab:L_BeB} and Table~\ref{tab:L_10Be9Be}). In the context of the analytical model, this behaviour can now be directly linked to the different ${\cal F}_{\rm HE}$ values in the two production sets, and more precisely for these sets, to the different production cross-sections of $^{10}$Be and $^9$Be from a few relevant progenitors ($^{12}$C, $^{16}$O, and $^{56}$Fe). The typical $20\%$ variation in ${\cal F}_{\rm HE}$ originating from these cross-section differences is responsible for a variation 
$$
(\Delta L/L)^{\mathrm{^{10}Be/^9Be}}_{\rm XS}\approx40\%,
$$
which is significantly larger than $(\Delta L/L)_{\rm CR~data}$.
We further discuss the uncertainties on ${\cal F}_{\rm HE}$ and comparisons with recent determination of $L$ in Appendix~\ref{app:comp}.

Finally, we note that the central values for $L$ in the Be/B or $\mathrm{^{10}Be/^9Be}$ analyses are similar for \xsGalxii{} (5~kpc vs. 4.7~kpc), but slightly different for \optxiiupxxii{} (3.8~kpc vs. 2.8~kpc). However, this difference is not significant and could easily be explained by uncertainties on the cross-sections for the production of $^{10}$B and $^{11}$B (involved in the Be/B analysis only).

\paragraph{Consequences for future experiments.}
The impact of production cross-sections for the determination of $L$ is significant, as has been already argued by different teams from Be/B analyses \citep{2020A&A...639A..74W,2020PhRvD.101b3013E,2021JCAP...03..099D}. As demonstrated here with $\mathrm{^{10}Be/^9Be}$, this dependence is directly tied in to the ${\cal F}_{\rm HE}$ term, showing that the value of the latter quantity is crucial for a good determination of $L$. Ideally, future experiments should go up to $\gtrsim100$~GeV/n in order to get a handle on this crucial ${\cal F}_{\rm HE}$ factor. However, AMS-02 and the HELIX project \citep{2019ICRC...36..121P} are expected to reach, at best, energies $\gtrsim 10$~GeV/n. This would already be a great experimental achievement but, unfortunately, these measurements may not be enough to strongly reduce the uncertainties on $L$. To fully benefit from the high-precision data of these experiments, the alternative is to improve the confidence (and reduce the scatter) that there is in the production cross-sections, in order to better constrain ${\cal F}(E_{k/n}),$ and thus $L$.

%_____________________________________________________________________________
%_____________________________________________________________________________
\section{Conclusions}
\label{sec:conclusions}

We have proposed a simple analytical formula to fit the halo size $L$ of the Galaxy from $(\mathrm{^{10}Be/^9Be})^{\rm TOA}$ data, without the need for an underlying propagation model. The minimal ingredients needed are: (i) the grammage (or rather the diffusion timescale), which can be directly taken from existing analyses of secondary-to-primary ratios; (ii) the destruction time, directly calculated from the inelastic cross-sections of $^{10}$Be and $^9$Be; and (iii) a high-energy normalisation constant ${\cal F}_{\rm HE}\approx 0.5$ (independent of $L$) or, for an even better accuracy, the energy-dependent ${\cal F}(E_{k/n})$ term calculated from inferred CR IS fluxes and nuclear data, and models for the production cross-sections of $^{10}$Be and $^9$Be.

We have also shown that the constraints set on $L$ from AMS-02 Be/B data using a combined fit and a propagation code are consistent with those obtained from a direct fit of existing $(\mathrm{^{10}Be/^9Be})^{\rm TOA}$ data with the analytical approximation. Moreover, with this approximation, uncertainties on the production cross-sections directly impact ${\cal F}_{\rm HE}$, and from various studies from the literature, we typically have $(\Delta{\cal F}_{\rm HE}/{\cal F}_{\rm HE})\approx 20\%$, which translates into $(\Delta L/L)\approx 40\%$. This uncertainty is clearly a limitation to fully exploiting forthcoming CR data. Fortunately, future data from the NA61/SHINE facility at the CERN SPS \citep{2019arXiv190907136U,2022icrc.confE.102A} should help to improve the situation. In particular, high-energy nuclear data for the production of Be isotopes from $^{16}$O (and to a lesser extent $^{56}$Fe) are desired.

It would be great if the analytical formula could be extended to other ratios of radioactive clocks. Unfortunately, this may prove difficult. First, the approximate formula is valid for $\mathrm{^{10}Be/^9Be}$, as long as it is expressed as a function of kinetic energy per nucleon (conserved quantity in fragmentation reactions); a similar formula for the ratio evaluated at the same rigidity is expected to be both more complicated and less accurate. Second, it would be nice to be able to extend the approximation for the Be/B ratio calculation. However, compared to the $\mathrm{^{10}Be/^9Be}$ case, we checked that a tentative formula for Be/B leads to larger biases, while the formula is to be used on a ratio that shows a weaker dependence on $L$ ($^{10}$Be is subdominant in Be). Third, one may think about applying the approximation to heavier CR clocks (e.g. $^{26}$Al, $^{36}$Cl, and $^{54}$Mn). However, the stable associated isotopes used to form the ratio of interest (e.g. $^{27}$Al in the $\mathrm{^{26}Al/^{27}Al}$ ratio) contain non-negligible primary source terms, which may prevent the derivation of a simple and accurate analytical formula.

Although the analytical model may only apply to the $\mathrm{^{10}Be/^9Be}$ ratio, we believe that it will already be very useful in order to interpret the forthcoming data from AMS-02 and HELIX. It should be also useful in order to quickly estimate the benefit that better production cross-sections of $^{10}$Be and $^{9}$Be can have on the determination of $L$.

%_____________________________________________________________________________
\begin{acknowledgements}
D.M. and L.D. thank their AMS colleagues for useful discussions that triggered the development of this project.
We thank Y. Génolini for his careful reading of the paper and comments.
We thank the Center for Information Technology of the University of Groningen for their support and for providing access to the Peregrine high-performance computing cluster.
This work was supported by the Programme National des Hautes Energies of CNRS/INSU with INP and IN2P3, co-funded by CEA and CNES.
\end{acknowledgements}

%_____________________________________________________________________________
\bibliographystyle{aa} % style aa.bst
\bibliography{10Be9Be}

\begin{thebibliography}{73}
\expandafter\ifx\csname natexlab\endcsname\relax\def\natexlab#1{#1}\fi

\bibitem[{{Aguilar} {et~al.}(2018a){Aguilar}, {Ali Cavasonza}, {Ambrosi},
  {Arruda}, {Attig}, {Aupetit}, {Azzarello}, {Bachlechner}, {Barao}, {Barrau},
  \& et~al.}]{2018PhRvL.120b1101A}
{Aguilar}, M., {Ali Cavasonza}, L., {Ambrosi}, G., {et~al.} 2018a, \prl, 120,
  021101

\bibitem[{{Aguilar} {et~al.}(2019){Aguilar}, {Ali Cavasonza}, {Ambrosi},
  {Arruda}, {Attig}, {Bachlechner}, {Barao}, {Barrau}, {Barrin}, {Bartoloni},
  {Basegmez-du Pree}, {Battiston}, {Becker}, {Behlmann}, {Beischer}, {Berdugo},
  {Bertucci}, {Bindi}, {de Boer}, {Bollweg}, {Borgia}, {Boschini}, {Bourquin},
  {Bueno}, {Burger}, {Burger}, {Cai}, {Capell}, {Caroff}, {Casaus},
  {Castellini}, {Cervelli}, {Chang}, {Chen}, {Chen}, {Chen}, {Cheng}, {Chou},
  {Choutko}, {Chung}, {Clark}, {Coignet}, {Consolandi}, {Contin}, {Corti},
  {Cui}, {Dadzie}, {Dai}, {Datta}, {Delgado}, {Della Torre}, {Demirk{\"o}z},
  {Derome}, {Di Falco}, {Di Felice}, {D{\'\i}az}, {Dimiccoli}, {von
  Doetinchem}, {Dong}, {Donnini}, {Duranti}, {Egorov}, {Eline}, {Feng}, {Fiand
  rini}, {Fisher}, {Formato}, {Galaktionov}, {G{\'a}mez},
  {Garc{\'\i}a-L{\'o}pez}, {Gargiulo}, {Gast}, {Gebauer}, {Gervasi},
  {Giovacchini}, {G{\'o}mez-Coral}, {Gong}, {Goy}, {Grabski}, {Grandi},
  {Graziani}, {Guo}, {Haino}, {Han}, {He}, {Hsieh}, {Huang}, {Huang},
  {Incagli}, {Jang}, {Jia}, {Jinchi}, {Kanishev}, {Khiali}, {Kim}, {Kirn},
  {Konyushikhin}, {Kounina}, {Kounine}, {Koutsenko}, {Kulemzin}, {La Vacca},
  {Laudi}, {Laurenti}, {Lazzizzera}, {Lebedev}, {Lee}, {Lee}, {Li}, {Li}, {Li},
  {Li}, {Light}, {Lin}, {Lippert}, {Liu}, {Lu}, {Lu}, {Luebelsmeyer}, {Luo},
  {Luo}, {Luo}, {Lyu}, {Machate}, {Ma{\~n}{\'a}}, {Mar{\'\i}n}, {Martin},
  {Mart{\'\i}nez}, {Masi}, {Maurin}, {Menchaca-Rocha}, {Meng}, {Mo}, {Molero},
  {Mott}, {Mussolin}, {Nelson}, {Ni}, {Nikonov}, {Nozzoli}, {Oliva}, {Orcinha},
  {Palermo}, {Palmonari}, {Paniccia}, {Pashnin}, {Pauluzzi}, {Pensotti},
  {Phan}, {Plyaskin}, {Poireau}, {Poluianov}, {Popkow}, {Qi}, {Qin}, {Qu},
  {Quadrani}, {Rancoita}, {Rapin}, {Reina Conde}, {Rosier-Lees}, {Rozhkov},
  {Rozza}, {Sagdeev}, {Schael}, {Schmidt}, {Schulz von Dratzig}, {Schwering},
  {Seo}, {Shan}, {Shi}, {Siedenburg}, {Solano}, {Song}, {Sun}, {Tacconi},
  {Tang}, {Tang}, {Tian}, {Ting}, {Ting}, {Tomassetti}, {Torsti},
  {T{\"u}ys{\"u}z}, {Urban}, {Usoskin}, {Vagelli}, {Vainio}, {Valente},
  {Valtonen}, {V{\'a}zquez Acosta}, {Vecchi}, {Velasco}, {Vialle}, {Wang},
  {Wang}, {Wang}, {Wang}, {Wang}, {Wang}, {Wei}, {Weng}, {Wu}, {Xiong}, {Xu},
  {Yan}, {Yang}, {Yi}, {Yu}, {Yu}, {Zannoni}, {Zeissler}, {Zhang}, {Zhang},
  {Zhang}, {Zhang}, {Zhao}, {Zheng}, {Zhuang}, {Zhukov}, {Zichichi},
  {Zimmermann}, {Zuccon}, \& {AMS Collaboration}}]{2019PhRvL.123r1102A}
{Aguilar}, M., {Ali Cavasonza}, L., {Ambrosi}, G., {et~al.} 2019, \prl, 123,
  181102

\bibitem[{{Aguilar} {et~al.}(2021{\natexlab{a}}){Aguilar}, {Ali Cavasonza},
  {Ambrosi}, {Arruda}, {Attig}, {Barao}, {Barrin}, {Bartoloni},
  {Ba{\c{s}}e{\u{g}}mez-du Pree}, {Bates}, {Battiston}, {Behlmann}, {Beischer},
  {Berdugo}, {Bertucci}, {Bindi}, {de Boer}, {Bollweg}, {Borgia}, {Boschini},
  {Bourquin}, {Bueno}, {Burger}, {Burger}, {Burmeister}, {Cai}, {Capell},
  {Casaus}, {Castellini}, {Cervelli}, {Chang}, {Chen}, {Chen}, {Chen}, {Cheng},
  {Chou}, {Chouridou}, {Choutko}, {Chung}, {Clark}, {Coignet}, {Consolandi},
  {Contin}, {Corti}, {Cui}, {Dadzie}, {Dai}, {Delgado}, {Della Torre},
  {Demirk{\"o}z}, {Derome}, {Di Falco}, {Di Felice}, {D{\'\i}az}, {Dimiccoli},
  {von Doetinchem}, {Dong}, {Donnini}, {Duranti}, {Egorov}, {Eline}, {Feng},
  {Fiandrini}, {Fisher}, {Formato}, {Freeman}, {Galaktionov}, {G{\'a}mez},
  {Garc{\'\i}a-L{\'o}pez}, {Gargiulo}, {Gast}, {Gebauer}, {Gervasi},
  {Giovacchini}, {G{\'o}mez-Coral}, {Gong}, {Goy}, {Grabski}, {Grandi},
  {Graziani}, {Guo}, {Haino}, {Han}, {Hashmani}, {He}, {Heber}, {Hsieh}, {Hu},
  {Huang}, {Hungerford}, {Incagli}, {Jang}, {Jia}, {Jinchi}, {Kanishev},
  {Khiali}, {Kim}, {Kirn}, {Konyushikhin}, {Kounina}, {Kounine}, {Koutsenko},
  {Kuhlman}, {Kulemzin}, {La Vacca}, {Laudi}, {Laurenti}, {Lazzizzera},
  {Lebedev}, {Lee}, {Lee}, {Leluc}, {Li}, {Li}, {Li}, {Li}, {Li}, {Li},
  {Light}, {Lin}, {Lippert}, {Liu}, {Lu}, {Lu}, {Luebelsmeyer}, {Luo}, {Lyu},
  {Machate}, {Ma{\~n}{\'a}}, {Mar{\'\i}n}, {Marquardt}, {Martin},
  {Mart{\'\i}nez}, {Masi}, {Maurin}, {Menchaca-Rocha}, {Meng}, {Mo}, {Molero},
  {Mott}, {Mussolin}, {Ni}, {Nikonov}, {Nozzoli}, {Oliva}, {Orcinha},
  {Palermo}, {Palmonari}, {Paniccia}, {Pashnin}, {Pauluzzi}, {Pensotti},
  {Phan}, {Plyaskin}, {Pohl}, {Porter}, {Qi}, {Qin}, {Qu}, {Quadrani},
  {Rancoita}, {Rapin}, {Reina Conde}, {Rosier-Lees}, {Rozhkov}, {Rozza},
  {Sagdeev}, {Schael}, {Schmidt}, {Schulz von Dratzig}, {Schwering}, {Seo},
  {Shan}, {Shi}, {Siedenburg}, {Solano}, {Song}, {Sonnabend}, {Sun}, {Sun},
  {Tacconi}, {Tang}, {Tang}, {Tian}, {Ting}, {Ting}, {Tomassetti}, {Torsti},
  {T{\"u}ys{\"u}z}, {Urban}, {Usoskin}, {Vagelli}, {Vainio}, {Valente},
  {Valtonen}, {V{\'a}zquez Acosta}, {Vecchi}, {Velasco}, {Vialle}, {Wang},
  {Wang}, {Wang}, {Wang}, {Wang}, {Wang}, {Wei}, {Weng}, {Wu}, {Xiong}, {Xu},
  {Yan}, {Yang}, {Yi}, {Yu}, {Yu}, {Zannoni}, {Zhang}, {Zhang}, {Zhang},
  {Zhang}, {Zhang}, {Zhao}, {Zheng}, {Zhuang}, {Zhukov}, {Zichichi},
  {Zimmermann}, {Zuccon}, \& {AMS Collaboration}}]{2021PhR...894....1A}
{Aguilar}, M., {Ali Cavasonza}, L., {Ambrosi}, G., {et~al.} 2021{\natexlab{a}},
  \physrep, 894, 1

\bibitem[{{Aguilar} {et~al.}(2021{\natexlab{b}}){Aguilar}, {Cavasonza},
  {Allen}, {Alpat}, {Ambrosi}, {Arruda}, {Attig}, {Barao}, {Barrin},
  {Bartoloni}, {Ba{\c{s}}e{\v{g}}mez-du Pree}, {Battiston}, {Behlmann},
  {Beischer}, {Berdugo}, {Bertucci}, {Bindi}, {de Boer}, {Bollweg}, {Borgia},
  {Boschini}, {Bourquin}, {Bueno}, {Burger}, {Burger}, {Burmeister}, {Cai},
  {Capell}, {Casaus}, {Castellini}, {Cervelli}, {Chang}, {Chen}, {Chen},
  {Chen}, {Chen}, {Cheng}, {Chou}, {Chouridou}, {Choutko}, {Chung}, {Clark},
  {Coignet}, {Consolandi}, {Contin}, {Corti}, {Cui}, {Dadzie}, {Delgado},
  {Della Torre}, {Demirk{\"o}z}, {Derome}, {Di Falco}, {Di Felice},
  {D{\'\i}az}, {Dimiccoli}, {von Doetinchem}, {Dong}, {Donnini}, {Duranti},
  {Egorov}, {Eline}, {Feng}, {Fiandrini}, {Fisher}, {Formato}, {Freeman},
  {Galaktionov}, {G{\'a}mez}, {Garc{\'\i}a-L{\'o}pez}, {Gargiulo}, {Gast},
  {Gervasi}, {Giovacchini}, {G{\'o}mez-Coral}, {Gong}, {Goy}, {Grabski},
  {Grandi}, {Graziani}, {Haino}, {Han}, {Hashmani}, {He}, {Heber}, {Hsieh},
  {Hu}, {Incagli}, {Jang}, {Jia}, {Jinchi}, {Kanishev}, {Khiali}, {Kim},
  {Kirn}, {Konyushikhin}, {Kounina}, {Kounine}, {Koutsenko}, {Kuhlman},
  {Kulemzin}, {La Vacca}, {Laudi}, {Laurenti}, {Lazzizzera}, {Lebedev}, {Lee},
  {Lee}, {Li}, {Li}, {Li}, {Li}, {Li}, {Li}, {Liang}, {Light}, {Lin},
  {Lippert}, {Liu}, {Liu}, {Lu}, {Lu}, {Luebelsmeyer}, {Luo}, {Luo}, {Lyu},
  {Machate}, {Ma{\~n}{\'a}}, {Mar{\'\i}n}, {Marquardt}, {Martin},
  {Mart{\'\i}nez}, {Masi}, {Maurin}, {Menchaca-Rocha}, {Meng}, {Mikhailov},
  {Mo}, {Molero}, {Mott}, {Mussolin}, {Negrete}, {Nikonov}, {Nozzoli}, {Oliva},
  {Orcinha}, {Palermo}, {Palmonari}, {Paniccia}, {Pashnin}, {Pauluzzi},
  {Pensotti}, {Phan}, {Piandani}, {Plyaskin}, {Poluianov}, {Qin}, {Qu},
  {Quadrani}, {Rancoita}, {Rapin}, {Conde}, {Robyn}, {Rosier-Lees}, {Rozhkov},
  {Rozza}, {Sagdeev}, {Schael}, {von Dratzig}, {Schwering}, {Seo}, {Shakfa},
  {Shan}, {Siedenburg}, {Solano}, {Song}, {Song}, {Sonnabend}, {Strigari},
  {Su}, {Sun}, {Sun}, {Tacconi}, {Tang}, {Tang}, {Tian}, {Ting}, {Ting},
  {Tomassetti}, {Torsti}, {T{\"u}ys{\"u}z}, {Urban}, {Usoskin}, {Vagelli},
  {Vainio}, {Valencia-Otero}, {Valente}, {Valtonen}, {V{\'a}zquez Acosta},
  {Vecchi}, {Velasco}, {Vialle}, {Wang}, {Wang}, {Wang}, {Wang}, {Wang},
  {Wang}, {Wang}, {Wang}, {Wang}, {Wei}, {Weng}, {Wu}, {Xiong}, {Xu}, {Yan},
  {Yang}, {Yashin}, {Yi}, {Yu}, {Yu}, {Zannoni}, {Zhang}, {Zhang}, {Zhang},
  {Zhang}, {Zhang}, {Zhao}, {Zheng}, {Zheng}, {Zhuang}, {Zhukov}, {Zichichi},
  {Zimmermann}, {Zuccon}, \& {AMS Collaboration}}]{2021PhRvL.126d1104A}
{Aguilar}, M., {Cavasonza}, L.~A., {Allen}, M.~S., {et~al.} 2021{\natexlab{b}},
  \prl, 126, 041104

\bibitem[{{Aguilar} {et~al.}(2021{\natexlab{c}}){Aguilar}, {Cavasonza},
  {Allen}, {Alpat}, {Ambrosi}, {Arruda}, {Attig}, {Barao}, {Barrin},
  {Bartoloni}, {Ba{\c{s}}e{\v{g}}mez-du Pree}, {Battiston}, {Behlmann},
  {Beranek}, {Berdugo}, {Bertucci}, {Bindi}, {Bollweg}, {Borgia}, {Boschini},
  {Bourquin}, {Bueno}, {Burger}, {Burger}, {Burmeister}, {Cai}, {Capell},
  {Casaus}, {Castellini}, {Cervelli}, {Chang}, {Chen}, {Chen}, {Chen}, {Chen},
  {Cheng}, {Chou}, {Chouridou}, {Choutko}, {Chung}, {Clark}, {Coignet},
  {Consolandi}, {Contin}, {Corti}, {Cui}, {Dadzie}, {Delgado}, {Della Torre},
  {Demirk{\"o}z}, {Derome}, {Di Falco}, {Di Felice}, {D{\'\i}az}, {Dimiccoli},
  {von Doetinchem}, {Dong}, {Donnini}, {Duranti}, {Egorov}, {Eline}, {Feng},
  {Fiandrini}, {Fisher}, {Formato}, {Freeman}, {Galaktionov}, {G{\'a}mez},
  {Garc{\'\i}a-L{\'o}pez}, {Gargiulo}, {Gast}, {Gervasi}, {Giovacchini},
  {G{\'o}mez-Coral}, {Gong}, {Goy}, {Grabski}, {Grandi}, {Graziani}, {Haino},
  {Han}, {Hashmani}, {He}, {Heber}, {Hsieh}, {Hu}, {Incagli}, {Jang}, {Jia},
  {Jinchi}, {Kanishev}, {Khiali}, {Kim}, {Kirn}, {Konyushikhin}, {Kounina},
  {Kounine}, {Koutsenko}, {Kuhlman}, {Kulemzin}, {La Vacca}, {Laudi},
  {Laurenti}, {Lazzizzera}, {Lebedev}, {Lee}, {Lee}, {Li}, {Li}, {Li}, {Li},
  {Li}, {Li}, {Liang}, {Light}, {Lin}, {Lippert}, {Liu}, {Liu}, {Lu}, {Lu},
  {Luebelsmeyer}, {Luo}, {Luo}, {Lyu}, {Machate}, {Ma{\~n}{\'a}}, {Mar{\'\i}n},
  {Marquardt}, {Martin}, {Mart{\'\i}nez}, {Masi}, {Maurin}, {Menchaca-Rocha},
  {Meng}, {Mikhailov}, {Mo}, {Molero}, {Mott}, {Mussolin}, {Negrete},
  {Nikonov}, {Nozzoli}, {Oliva}, {Orcinha}, {Palermo}, {Palmonari}, {Paniccia},
  {Pashnin}, {Pauluzzi}, {Pensotti}, {Phan}, {Piandani}, {Plyaskin},
  {Poluianov}, {Qin}, {Qu}, {Quadrani}, {Rancoita}, {Rapin}, {Conde}, {Robyn},
  {Rosier-Lees}, {Rozhkov}, {Rozza}, {Sagdeev}, {Schael}, {Schulz von Dratzig},
  {Schwering}, {Seo}, {Shakfa}, {Shan}, {Siedenburg}, {Solano}, {Song}, {Song},
  {Sonnabend}, {Strigari}, {Su}, {Sun}, {Sun}, {Tacconi}, {Tang}, {Tang},
  {Tian}, {Ting}, {Ting}, {Tomassetti}, {Torsti}, {T{\"u}ys{\"u}z}, {Urban},
  {Usoskin}, {Vagelli}, {Vainio}, {Valencia-Otero}, {Valente}, {Valtonen},
  {V{\'a}zquez Acosta}, {Vecchi}, {Velasco}, {Vialle}, {Wang}, {Wang}, {Wang},
  {Wang}, {Wang}, {Wang}, {Wang}, {Wang}, {Wang}, {Wei}, {Weng}, {Wu}, {Xiong},
  {Xu}, {Yan}, {Yang}, {Yashin}, {Yi}, {Yu}, {Yu}, {Zannoni}, {Zhang}, {Zhang},
  {Zhang}, {Zhang}, {Zhang}, {Zhao}, {Zheng}, {Zheng}, {Zhuang}, {Zhukov},
  {Zichichi}, {Zuccon}, \& {AMS Collaboration}}]{2021PhRvL.126h1102A}
{Aguilar}, M., {Cavasonza}, L.~A., {Allen}, M.~S., {et~al.} 2021{\natexlab{c}},
  \prl, 126, 081102

\bibitem[{{Aguilar} {et~al.}(2021{\natexlab{d}}){Aguilar}, {Cavasonza},
  {Alpat}, {Ambrosi}, {Arruda}, {Attig}, {Barao}, {Barrin}, {Bartoloni},
  {Ba{\c{s}}e{\v{g}}mez-du Pree}, {Battiston}, {Behlmann}, {Beranek},
  {Berdugo}, {Bertucci}, {Bindi}, {Bollweg}, {Borgia}, {Boschini}, {Bourquin},
  {Bueno}, {Burger}, {Burger}, {Burmeister}, {Cai}, {Capell}, {Casaus},
  {Castellini}, {Cervelli}, {Chang}, {Chen}, {Chen}, {Chen}, {Chen}, {Cheng},
  {Chou}, {Chouridou}, {Choutko}, {Chung}, {Clark}, {Coignet}, {Consolandi},
  {Contin}, {Corti}, {Cui}, {Dadzie}, {Delgado}, {Della Torre}, {Demirk{\"o}z},
  {Derome}, {Di Falco}, {Di Felice}, {D{\'\i}az}, {Dimiccoli}, {von
  Doetinchem}, {Dong}, {Donnini}, {Duranti}, {Egorov}, {Eline}, {Feng},
  {Fiandrini}, {Fisher}, {Formato}, {Freeman}, {G{\'a}mez},
  {Garc{\'\i}a-L{\'o}pez}, {Gargiulo}, {Gast}, {Gervasi}, {Giovacchini},
  {G{\'o}mez-Coral}, {Gong}, {Goy}, {Grabski}, {Grandi}, {Graziani}, {Haino},
  {Han}, {Hashmani}, {He}, {Heber}, {Hsieh}, {Hu}, {Incagli}, {Jang}, {Jia},
  {Jinchi}, {Khiali}, {Kim}, {Kirn}, {Konyushikhin}, {Kounina}, {Kounine},
  {Koutsenko}, {Krasnopevtsev}, {Kuhlman}, {Kulemzin}, {La Vacca}, {Laudi},
  {Laurenti}, {Lazzizzera}, {Lebedev}, {Lee}, {Lee}, {Li}, {Li}, {Li}, {Li},
  {Li}, {Li}, {Liang}, {Light}, {Lin}, {Lippert}, {Liu}, {Liu}, {Lu}, {Lu},
  {Luebelsmeyer}, {Luo}, {Luo}, {Machate}, {Ma{\~n}{\'a}}, {Mar{\'\i}n},
  {Marquardt}, {Martin}, {Mart{\'\i}nez}, {Masi}, {Maurin}, {Medvedeva},
  {Menchaca-Rocha}, {Meng}, {Mikhailov}, {Molero}, {Mott}, {Mussolin},
  {Negrete}, {Nikonov}, {Nozzoli}, {Oliva}, {Orcinha}, {Palermo}, {Palmonari},
  {Paniccia}, {Pashnin}, {Pauluzzi}, {Pensotti}, {Phan}, {Plyaskin}, {Pohl},
  {Poluianov}, {Qin}, {Qu}, {Quadrani}, {Rancoita}, {Rapin}, {Conde}, {Robyn},
  {Rosier-Lees}, {Rozhkov}, {Rozza}, {Sagdeev}, {Schael}, {von Dratzig},
  {Schwering}, {Seo}, {Shakfa}, {Shan}, {Siedenburg}, {Solano}, {Song}, {Song},
  {Sonnabend}, {Strigari}, {Su}, {Sun}, {Sun}, {Tacconi}, {Tang}, {Tang},
  {Tian}, {Ting}, {Ting}, {Tomassetti}, {Torsti}, {T{\"u}ys{\"u}z}, {Urban},
  {Usoskin}, {Vagelli}, {Vainio}, {Valencia-Otero}, {Valente}, {Valtonen},
  {V{\'a}zquez Acosta}, {Vecchi}, {Velasco}, {Vialle}, {Wang}, {Wang}, {Wang},
  {Wang}, {Wang}, {Wang}, {Wang}, {Wang}, {Wang}, {Wei}, {Weng}, {Wu}, {Xiong},
  {Xu}, {Yan}, {Yang}, {Yashin}, {Yi}, {Yu}, {Yu}, {Zannoni}, {Zhang}, {Zhang},
  {Zhang}, {Zhang}, {Zhang}, {Zhao}, {Zheng}, {Zheng}, {Zhuang}, {Zhukov},
  {Zichichi}, {Zuccon}, \& {AMS Collaboration}}]{2021PhRvL.127b1101A}
{Aguilar}, M., {Cavasonza}, L.~A., {Alpat}, B., {et~al.} 2021{\natexlab{d}},
  \prl, 127, 021101

\bibitem[{{Amin} \& {NA61/SHINE}(2021)}]{2022icrc.confE.102A}
{Amin}, N. \& {NA61/SHINE}. 2021, in \icrc 37, 102

\bibitem[{{Boschini} {et~al.}(2020{\natexlab{a}}){Boschini}, {Della Torre},
  {Gervasi}, {Grandi}, {J{\'o}hannesson}, {La Vacca}, {Masi}, {Moskalenko},
  {Pensotti}, {Porter}, {Quadrani}, {Rancoita}, {Rozza}, \&
  {Tacconi}}]{2020ApJS..250...27B}
{Boschini}, M.~J., {Della Torre}, S., {Gervasi}, M., {et~al.}
  2020{\natexlab{a}}, \apjs, 250, 27

\bibitem[{{Boschini} {et~al.}(2020{\natexlab{b}}){Boschini}, {Torre},
  {Gervasi}, {Grand i}, {J{\o}hannesson}, {Vacca}, {Masi}, {Moskalenko},
  {Pensotti}, {Porter}, {Quadrani}, {Rancoita}, {Rozza}, \&
  {Tacconi}}]{2020ApJ...889..167B}
{Boschini}, M.~J., {Torre}, S.~D., {Gervasi}, M., {et~al.} 2020{\natexlab{b}},
  \apj, 889, 167

\bibitem[{{Boudaud} {et~al.}(2020){Boudaud}, {G{\'e}nolini}, {Derome},
  {Lavalle}, {Maurin}, {Salati}, \& {Serpico}}]{2020PhRvR...2b3022B}
{Boudaud}, M., {G{\'e}nolini}, Y., {Derome}, L., {et~al.} 2020, \prr, 2, 023022

\bibitem[{{Connell}(1998)}]{1998ApJ...501L..59C}
{Connell}, J.~J. 1998, \apjl, 501, L59

\bibitem[{{Cummings} {et~al.}(2016){Cummings}, {Stone}, {Heikkila}, {Lal},
  {Webber}, {J{\'o}hannesson}, {Moskalenko}, {Orlando}, \&
  {Porter}}]{2016ApJ...831...18C}
{Cummings}, A.~C., {Stone}, E.~C., {Heikkila}, B.~C., {et~al.} 2016, \apj, 831,
  18

\bibitem[{{De La Torre Luque} {et~al.}(2022){De La Torre Luque}, {Mazziotta},
  {Ferrari}, {Loparco}, {Sala}, \& {Serini}}]{2022JCAP...07..008D}
{De La Torre Luque}, P., {Mazziotta}, M.~N., {Ferrari}, A., {et~al.} 2022,
  \jcap, 2022, 008

\bibitem[{{De La Torre Luque} {et~al.}(2021){De La Torre Luque}, {Mazziotta},
  {Loparco}, {Gargano}, \& {Serini}}]{2021JCAP...03..099D}
{De La Torre Luque}, P., {Mazziotta}, M.~N., {Loparco}, F., {Gargano}, F., \&
  {Serini}, D. 2021, \jcap, 2021, 099

\bibitem[{{Derome} {et~al.}(2019){Derome}, {Maurin}, {Salati}, {Boudaud},
  {G{\'e}nolini}, \& {Kunz{\'e}}}]{2019A&A...627A.158D}
{Derome}, L., {Maurin}, D., {Salati}, P., {et~al.} 2019, \aap, 627, A158

\bibitem[{{Di Bernardo} {et~al.}(2010){Di Bernardo}, {Evoli}, {Gaggero},
  {Grasso}, \& {Maccione}}]{2010APh....34..274D}
{Di Bernardo}, G., {Evoli}, C., {Gaggero}, D., {Grasso}, D., \& {Maccione}, L.
  2010, Astropart. Phys., 34, 274

\bibitem[{{Donato} {et~al.}(2004){Donato}, {Fornengo}, {Maurin}, {Salati}, \&
  {Taillet}}]{Donato2004}
{Donato}, F., {Fornengo}, N., {Maurin}, D., {Salati}, P., \& {Taillet}, R.
  2004, \prd, 69, 063501

\bibitem[{{Donato} {et~al.}(2002){Donato}, {Maurin}, \&
  {Taillet}}]{2002A&A...381..539D}
{Donato}, F., {Maurin}, D., \& {Taillet}, R. 2002, \aap, 381, 539

\bibitem[{{Engelmann} {et~al.}(1990){Engelmann}, {Ferrando}, {Soutoul},
  {Goret}, \& {Juliusson}}]{1990A&A...233...96E}
{Engelmann}, J.~J., {Ferrando}, P., {Soutoul}, A., {Goret}, P., \& {Juliusson},
  E. 1990, \aap, 233, 96

\bibitem[{{Evoli} {et~al.}(2019){Evoli}, {Aloisio}, \&
  {Blasi}}]{2019PhRvD..99j3023E}
{Evoli}, C., {Aloisio}, R., \& {Blasi}, P. 2019, \prd, 99, 103023

\bibitem[{{Evoli} {et~al.}(2018){Evoli}, {Gaggero}, {Vittino}, {Di Mauro},
  {Grasso}, \& {Mazziotta}}]{2018JCAP...07..006E}
{Evoli}, C., {Gaggero}, D., {Vittino}, A., {et~al.} 2018, \jcap, 2018, 006

\bibitem[{{Evoli} {et~al.}(2020){Evoli}, {Morlino}, {Blasi}, \&
  {Aloisio}}]{2020PhRvD.101b3013E}
{Evoli}, C., {Morlino}, G., {Blasi}, P., \& {Aloisio}, R. 2020, \prd, 101,
  023013

\bibitem[{{Ferri{\`e}re}(2001)}]{2001RvMP...73.1031F}
{Ferri{\`e}re}, K.~M. 2001, Reviews of Modern Physics, 73, 1031

\bibitem[{{Garcia-Munoz} {et~al.}(1981){Garcia-Munoz}, {Simpson}, \&
  {Wefel}}]{1981ICRC....2...72G}
{Garcia-Munoz}, M., {Simpson}, J.~A., \& {Wefel}, J.~P. 1981, in \icrc 17,
  Vol.~2, 72--75

\bibitem[{{G{\'e}nolini} {et~al.}(2019){G{\'e}nolini}, {Boudaud}, {Batista},
  {Caroff}, {Derome}, {Lavalle}, {Marcowith}, {Maurin}, {Poireau}, {Poulin},
  {Rosier}, {Salati}, {Serpico}, \& {Vecchi}}]{2019PhRvD..99l3028G}
{G{\'e}nolini}, Y., {Boudaud}, M., {Batista}, P.~I., {et~al.} 2019, \prd, 99,
  123028

\bibitem[{{G{\'e}nolini} {et~al.}(2021){G{\'e}nolini}, {Boudaud}, {Cirelli},
  {Derome}, {Lavalle}, {Maurin}, {Salati}, \& {Weinrich}}]{2021PhRvD.104h3005G}
{G{\'e}nolini}, Y., {Boudaud}, M., {Cirelli}, M., {et~al.} 2021, \prd, 104,
  083005

\bibitem[{{G{\'e}nolini} {et~al.}(2018){G{\'e}nolini}, {Maurin}, {Moskalenko},
  \& {Unger}}]{2018PhRvC..98c4611G}
{G{\'e}nolini}, Y., {Maurin}, D., {Moskalenko}, I.~V., \& {Unger}, M. 2018,
  \prc, 98, 034611

\bibitem[{{George} {et~al.}(2009){George}, {Lave}, {Wiedenbeck}, {Binns},
  {Cummings}, {Davis}, {de Nolfo}, {Hink}, {Israel}, {Leske}, {Mewaldt},
  {Scott}, {Stone}, {von Rosenvinge}, \& {Yanasak}}]{2009ApJ...698.1666G}
{George}, J.~S., {Lave}, K.~A., {Wiedenbeck}, M.~E., {et~al.} 2009, \apj, 698,
  1666

\bibitem[{{Ghelfi} {et~al.}(2016){Ghelfi}, {Barao}, {Derome}, \&
  {Maurin}}]{2016A&A...591A..94G}
{Ghelfi}, A., {Barao}, F., {Derome}, L., \& {Maurin}, D. 2016, \aap, 591, A94

\bibitem[{{Ghelfi} {et~al.}(2017){Ghelfi}, {Maurin}, {Cheminet}, {Derome},
  {Hubert}, \& {Melot}}]{2017AdSpR..60..833G}
{Ghelfi}, A., {Maurin}, D., {Cheminet}, A., {et~al.} 2017, AdSR, 60, 833

\bibitem[{{Ginzburg} {et~al.}(1980){Ginzburg}, {Khazan}, \&
  {Ptuskin}}]{1980Ap&SS..68..295G}
{Ginzburg}, V.~L., {Khazan}, I.~M., \& {Ptuskin}, V.~S. 1980, Astrophys. Space
  Sci., 68, 295

\bibitem[{{Gleeson} \& {Axford}(1967)}]{1967ApJ...149L.115G}
{Gleeson}, L.~J. \& {Axford}, W.~I. 1967, \apjl, 149, L115

\bibitem[{{Gleeson} \& {Axford}(1968)}]{1968ApJ...154.1011G}
{Gleeson}, L.~J. \& {Axford}, W.~I. 1968, \apj, 154, 1011

\bibitem[{{Hams} {et~al.}(2004){Hams}, {Barbier}, {Bremerich}, {Christian}, {de
  Nolfo}, {Geier}, {G{\"o}bel}, {Gupta}, {Hof}, {Menn}, {Mewaldt}, {Mitchell},
  {Schindler}, {Simon}, \& {Streitmatter}}]{2004ApJ...611..892H}
{Hams}, T., {Barbier}, L.~M., {Bremerich}, M., {et~al.} 2004, \apj, 611, 892

\bibitem[{{Hayakawa} {et~al.}(1958){Hayakawa}, {Ito}, \&
  {Terashima}}]{1958PThPS...6....1H}
{Hayakawa}, S., {Ito}, K., \& {Terashima}, Y. 1958, Prog. Theor. Phys. Supp.,
  6, 1

\bibitem[{{Jones} {et~al.}(2001){Jones}, {Lukasiak}, {Ptuskin}, \&
  {Webber}}]{2001ApJ...547..264J}
{Jones}, F.~C., {Lukasiak}, A., {Ptuskin}, V., \& {Webber}, W. 2001, \apj, 547,
  264

\bibitem[{{Korsmeier} \& {Cuoco}(2021)}]{2021PhRvD.103j3016K}
{Korsmeier}, M. \& {Cuoco}, A. 2021, \prd, 103, 103016

\bibitem[{{Lave} {et~al.}(2013){Lave}, {Wiedenbeck}, {Binns}, {Christian},
  {Cummings}, {Davis}, {de Nolfo}, {Israel}, {Leske}, {Mewaldt}, {Stone}, \&
  {von Rosenvinge}}]{2013ApJ...770..117L}
{Lave}, K.~A., {Wiedenbeck}, M.~E., {Binns}, W.~R., {et~al.} 2013, \apj, 770,
  117

\bibitem[{{Lukasiak}(1999)}]{1999ICRC....3...41L}
{Lukasiak}, A. 1999, in \icrc 26, Vol.~3, 41

\bibitem[{{Maurin}(2020)}]{2020CoPhC.24706942M}
{Maurin}, D. 2020, \cpc, 247, 106942

\bibitem[{{Maurin} {et~al.}(2015){Maurin}, {Cheminet}, {Derome}, {Ghelfi}, \&
  {Hubert}}]{2015AdSpR..55..363M}
{Maurin}, D., {Cheminet}, A., {Derome}, L., {Ghelfi}, A., \& {Hubert}, G. 2015,
  AdSR, 55, 363

\bibitem[{{Maurin} {et~al.}(2020){Maurin}, {Dembinski}, {Gonzalez},
  {Mari{\c{s}}}, \& {Melot}}]{2020Univ....6..102M}
{Maurin}, D., {Dembinski}, H.~P., {Gonzalez}, J., {Mari{\c{s}}}, I.~C., \&
  {Melot}, F. 2020, Universe, 6, 102

\bibitem[{{Maurin} {et~al.}(2001){Maurin}, {Donato}, {Taillet}, \&
  {Salati}}]{2001ApJ...555..585M}
{Maurin}, D., {Donato}, F., {Taillet}, R., \& {Salati}, P. 2001, \apj, 555, 585

\bibitem[{{Maurin} {et~al.}(2022){Maurin}, {Ferronato Bueno}, {G{\'e}nolini},
  {Derome}, \& {Vecchi}}]{2022arXiv220300522M}
{Maurin}, D., {Ferronato Bueno}, E., {G{\'e}nolini}, Y., {Derome}, L., \&
  {Vecchi}, M. 2022, \aap\ (accepted), arXiv:2203.00522

\bibitem[{{Maurin} {et~al.}(2014){Maurin}, {Melot}, \&
  {Taillet}}]{2014A&A...569A..32M}
{Maurin}, D., {Melot}, F., \& {Taillet}, R. 2014, \aap, 569, A32

\bibitem[{{Maurin} {et~al.}(2010){Maurin}, {Putze}, \&
  {Derome}}]{2010A&A...516A..67M}
{Maurin}, D., {Putze}, A., \& {Derome}, L. 2010, \aap, 516, A67

\bibitem[{{Moskalenko} {et~al.}(2001){Moskalenko}, {Mashnik}, \&
  {Strong}}]{2001ICRC....5.1836M}
{Moskalenko}, I.~V., {Mashnik}, S.~G., \& {Strong}, A.~W. 2001, in , 1836--1839

\bibitem[{{Nozzoli} \& {Cernetti}(2021)}]{2021Univ....7..183N}
{Nozzoli}, F. \& {Cernetti}, C. 2021, Universe, 7, 183

\bibitem[{{O'dell} {et~al.}(1975){O'dell}, {Shapiro}, {Silberberg}, \&
  {Tsao}}]{1975ICRC....2..526O}
{O'dell}, F.~W., {Shapiro}, M.~M., {Silberberg}, R., \& {Tsao}, C.~H. 1975, in
  \icrc 14, Vol.~2, 526

\bibitem[{{Park} {et~al.}(2019){Park}, {Beaufore}, {Mbarek}, {Muller},
  {Schreyer}, {Wakely}, {Werner}, {Wisher}, {Tabata}, {Gebhard}, {Kunkler},
  {Musser}, {Michaels}, {Visser}, {Ellingwood}, {Hanna}, {O'Brien}, {Rosin},
  {Nutter}, {Allison}, {Beatty}, {McBride}, {Chen}, {Coutu}, {Mognet}, {Yu},
  {Green}, {Tarle}, \& {Tomasch}}]{2019ICRC...36..121P}
{Park}, N., {Beaufore}, L., {Mbarek}, R., {et~al.} 2019, in \icrc 36, Vol.~36,
  121

\bibitem[{{Porter} {et~al.}(2021){Porter}, {Johannesson}, \&
  {Moskalenko}}]{2021arXiv211212745P}
{Porter}, T.~A., {Johannesson}, G., \& {Moskalenko}, I.~V. 2021,
  arXiv:2112.12745

\bibitem[{{Potgieter}(2013)}]{2013LRSP...10....3P}
{Potgieter}, M. 2013, Living Rev. Sol. Phys., 10, 3

\bibitem[{{Prishchep} \& {Ptuskin}(1975)}]{1975Ap&SS..32..265P}
{Prishchep}, V.~L. \& {Ptuskin}, V.~S. 1975, \apss, 32, 265

\bibitem[{{Putze} {et~al.}(2010){Putze}, {Derome}, \&
  {Maurin}}]{2010A&A...516A..66P}
{Putze}, A., {Derome}, L., \& {Maurin}, D. 2010, \aap, 516, A66

\bibitem[{{Putze} {et~al.}(2011){Putze}, {Maurin}, \&
  {Donato}}]{2011A&A...526A.101P}
{Putze}, A., {Maurin}, D., \& {Donato}, F. 2011, \aap, 526, A101

\bibitem[{{Reinert} \& {Winkler}(2018)}]{2018JCAP...01..055R}
{Reinert}, A. \& {Winkler}, M.~W. 2018, \jcap, 2018, 055

\bibitem[{{Schroer} {et~al.}(2021){Schroer}, {Evoli}, \&
  {Blasi}}]{2021PhRvD.103l3010S}
{Schroer}, B., {Evoli}, C., \& {Blasi}, P. 2021, \prd, 103, 123010

\bibitem[{{Shen} {et~al.}(2019){Shen}, {Qin}, {Zuo}, \&
  {Wei}}]{2019ApJ...887..132S}
{Shen}, Z.~N., {Qin}, G., {Zuo}, P., \& {Wei}, F. 2019, \apj, 887, 132

\bibitem[{{Silberberg} \& {Tsao}(1990)}]{1990PhR...191..351S}
{Silberberg}, R. \& {Tsao}, C.~H. 1990, \physrep, 191, 351

\bibitem[{{Simpson} \& {Garcia-Munoz}(1988)}]{1988SSRv...46..205S}
{Simpson}, J.~A. \& {Garcia-Munoz}, M. 1988, \ssr, 46, 205

\bibitem[{{Strong} {et~al.}(2007){Strong}, {Moskalenko}, \&
  {Ptuskin}}]{Strong2007}
{Strong}, A.~W., {Moskalenko}, I.~V., \& {Ptuskin}, V.~S. 2007, Annu. Rev.
  Nucl. Part. Sci., 57, 285

\bibitem[{{Tripathi} {et~al.}(1997){Tripathi}, {Cucinotta}, \&
  {Wilson}}]{1997lrc..reptQ....T}
{Tripathi}, R.~K., {Cucinotta}, F.~A., \& {Wilson}, J.~W. 1997, {Universal
  Parameterization of Absorption Cross Sections}, Tech. rep., NASA Langley
  Research Center

\bibitem[{{Tripathi} {et~al.}(1999){Tripathi}, {Cucinotta}, \&
  {Wilson}}]{1999STIN...0004259T}
{Tripathi}, R.~K., {Cucinotta}, F.~A., \& {Wilson}, J.~W. 1999, {Universal
  Parameterization of Absorption Cross Sections - Light systems}, Tech. rep.,
  NASA Langley Research Center

\bibitem[{{Unger} \& NA61/SHINE(2019)}]{2019arXiv190907136U}
{Unger}, M. \& NA61/SHINE. 2019, in \icrc 36, 446

\bibitem[{{Vecchi} {et~al.}(2022){Vecchi}, {Batista}, {Bueno}, {Derome},
  {G{\'e}nolini}, \& {Maurin}}]{2022arXiv220306479V}
{Vecchi}, M., {Batista}, P.~I., {Bueno}, E.~F., {et~al.} 2022, Front. Phys.,
  10, 858841

\bibitem[{{Wang} {et~al.}(2021){Wang}, {Wu}, \& {Long}}]{2021arXiv210803687W}
{Wang}, Y., {Wu}, J., \& {Long}, W.-C. 2021, arXiv e-prints, arXiv:2108.03687

\bibitem[{{Webber} \& {Soutoul}(1998)}]{1998ApJ...506..335W}
{Webber}, W.~R. \& {Soutoul}, A. 1998, \apj, 506, 335

\bibitem[{{Webber} {et~al.}(2003){Webber}, {Soutoul}, {Kish}, \&
  {Rockstroh}}]{2003ApJS..144..153W}
{Webber}, W.~R., {Soutoul}, A., {Kish}, J.~C., \& {Rockstroh}, J.~M. 2003,
  \apjs, 144, 153

\bibitem[{{Weinrich} {et~al.}(2020{\natexlab{a}}){Weinrich}, {Boudaud},
  {Derome}, {G{\'e}nolini}, {Lavalle}, {Maurin}, {Salati}, {Serpico}, \&
  {Weymann-Despres}}]{2020A&A...639A..74W}
{Weinrich}, N., {Boudaud}, M., {Derome}, L., {et~al.} 2020{\natexlab{a}}, \aap,
  639, A74

\bibitem[{{Weinrich} {et~al.}(2020{\natexlab{b}}){Weinrich}, {G{\'e}nolini},
  {Boudaud}, {Derome}, \& {Maurin}}]{2020A&A...639A.131W}
{Weinrich}, N., {G{\'e}nolini}, Y., {Boudaud}, M., {Derome}, L., \& {Maurin},
  D. 2020{\natexlab{b}}, \aap, 639, A131

\bibitem[{{Wiedenbeck} \& {Greiner}(1980)}]{1980ApJ...239L.139W}
{Wiedenbeck}, M.~E. \& {Greiner}, D.~E. 1980, \apjl, 239, L139

\bibitem[{{Yanasak} {et~al.}(2001){Yanasak}, {Wiedenbeck}, {Mewaldt}, {Davis},
  {Cummings}, {George}, {Leske}, {Stone}, {Christian}, {von Rosenvinge},
  {Binns}, {Hink}, \& {Israel}}]{2001ApJ...563..768Y}
{Yanasak}, N.~E., {Wiedenbeck}, M.~E., {Mewaldt}, R.~A., {et~al.} 2001, \apj,
  563, 768

\bibitem[{{Yuan} {et~al.}(2020){Yuan}, {Zhu}, {Bi}, \&
  {Wei}}]{2020JCAP...11..027Y}
{Yuan}, Q., {Zhu}, C.-R., {Bi}, X.-J., \& {Wei}, D.-M. 2020, \jcap, 2020, 027

\end{thebibliography}

%_____________________________________________________________________________
%_____________________________________________________________________________
\clearpage
\appendix
\section{Comparison with other studies and the impact of re-acceleration}
\label{app:comp}

For comparison purposes, we tried to repeat our analysis using ${\cal F}_{\rm HE}$ and $t_{\rm diff}$ values from other publications. The only recent work for which we could retrieve the necessary ingredients is that of \citet{2021JCAP...03..099D}, hereafter [De21]. In the study, the authors compare the use of different production cross-section sets for the determination of the halo size $L$.

\subsection{${\cal F}_{\rm HE}$ from various cross-section sets}
\label{app:XS}
Taking the asymptotic high-energy value of $\mathrm{^{10}Be/^9Be}$ from Fig.~8 of [De21] and using Eq.~(\ref{eq:1Ddiff}), we can calculate ${\cal F}_{\rm HE}$ associated with their cross-section sets ({\tt Webber}, {\tt GALPROP}, and {\tt DRAGON2}), with ${\cal F}_{\rm HE}=0.35$, 0.60, and 0.58 respectively.

The cross-sections sets used in [De21] were introduced, and discussed in detail, in \citet{2018JCAP...07..006E}, and we can compare them to those used in our analysis. The {\tt GALPROP} set in [De21] and \xsGalxii{} used here are one and the same. Although the underlying propagation model (and to some extent, the CR data used) are different in these two studies, we obtain very similar ${\cal F}_{\rm HE}$ values (${\cal F}_{\rm HE}^{\tt GALPROP}=0.60$ vs. ${\cal F}_{\rm HE}^\xsGalxii{}=0.62$ reported in our Table~\ref{tab:L_10Be9Be}); also similar to the value obtained for {\tt DRAGON2} (${\cal F}_{\rm HE}=0.58$)\footnote{We cannot directly compare with the value obtained in Fig.~5 of \citet{2020PhRvD.101b3013E} as the latter is shown as a function of rigidity and not the kinetic energy per nucleon.}. These values are also similar to the one obtained directly from the \galprop{} team (using their {\tt GALPROP} cross-section set), as can be calculated from the high-energy $\mathrm{^{10}Be/^9Be}$ ratio shown in Fig.~11 of \citet{Strong2007}.

However, the very low value ${\cal F}_{\rm HE}^{\tt WEBBER}=0.35$ in [De21] is quite puzzling. As described in \citet{2018JCAP...07..006E}, the so-called {\tt Webber} set is a mixture of Webber's 2003 cross-section values \citep{2003ApJS..144..153W}, hereafter {\tt W03}, and calculations based on the Webber's 1983 values. The motivation for this hybrid set is not completely straightforward, as {\tt W03} should completely supersede  Webber's older calculations. The {\tt W03} set was used in \citet{2010A&A...516A..66P}\footnote{We can track, without ambiguity, that the {\tt W03} set that both studies refer to is the same, as acknowledged in \citet{2010APh....34..274D}.}, where their Fig.~10 allows ${\cal F}_{\rm HE}=0.58$ to be calculated, a value actually in line with that of the other cross-section sets. We therefore recommend caution when using this hybrid {\tt Webber} set.

In any case, this comparison shows that the various production cross-section sets used in the literature agree within $10-20\%$. Unfortunately, as highlighted in the text, a 20\% difference translates into a $\sim 40\%$ uncertainty on $L$.

\subsection{Comparison with $L$ values from [De21]}

We can try to reproduce the $L$ values obtained in [De21] from our analytical formula, extracting $t_{\rm diff}$ from their Eq.~(2.2) and Table 2.3\footnote{The propagation model in [De21] has a more realistic spatial distribution of the gas than ours, but we nevertheless assume $h=0.1$~kpc to calculate $t_{\rm diff}$ from their diffusion coefficient: this is the typical scale height of the gas density, and the value allowing the recovery of the correct gas column density \citep{2001RvMP...73.1031F}}.
Using the same data as in [De21], we find $L_{\tt DRAGON2}=3.4$~kpc and $L_{\tt WEBBER}=1.1$~kpc, to compare to their publication values $6.8$~kpc and $2.1$~kpc respectively. Despite these significant differences, the strong correlation between $L$ and ${\cal F}_{\rm HE}$ remains, and confirms that the uncertainty on ${\cal F}_{\rm HE}$ is the dominant source of uncertainty for the determination of $L$.
In addition, we would have expected our analytical formula to lead to similar values for $L_{\tt DRAGON2}$ and $L_{\optxiiupxxii{}}$, as ${\cal F}_{\rm HE}(\optxiiupxxii{})\approx{\cal F}_{\rm HE}({\tt DRAGON2})$, but we only get $L_{\tt DRAGON2}=3.4$~kpc and $L_{\optxiiupxxii{}}=2.8$~kpc.

\begin{figure}[!t]
   \centering
   \includegraphics[width=\columnwidth]{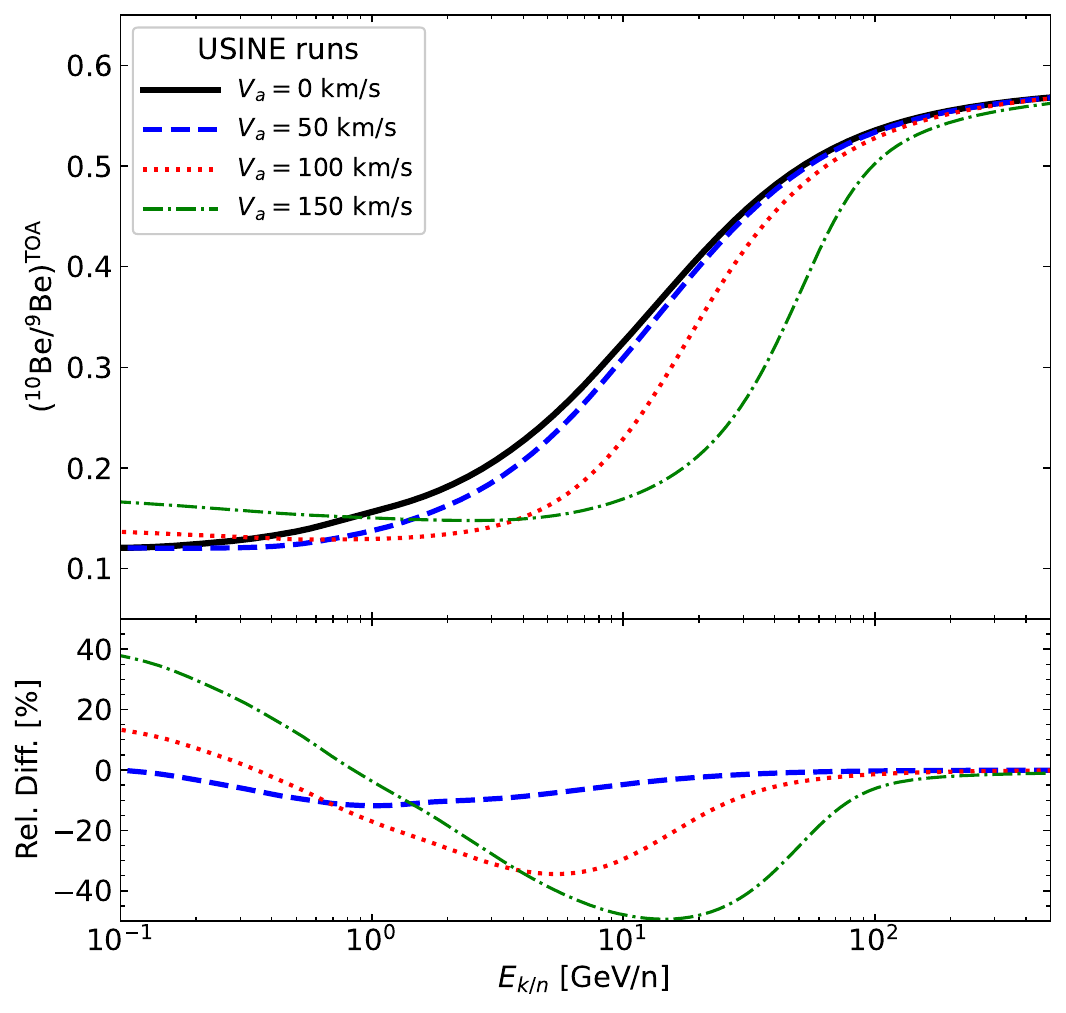}
   \caption{$\mathrm{^{10}Be/^9Be}$ TOA ratio (top) with or without re-acceleration (in the thin disc only, see text) and relative difference with respect to the $V_a=0$ case (bottom).
   \label{fig:impact_Va}}
\end{figure}

The likely origin of the above differences is the presence of re-acceleration in [De21]---we checked that varying $h$ or the subset of data used in the fit is not enough to explain them. To test this hypothesis, we calculate, with the \usine{} code, the $\mathrm{^{10}Be/^9Be}$ ratio, varying the re-acceleration, but leaving unchanged the diffusion time.
We stress that in the model used in [De21], the re-acceleration, mediated by the Alfv\'enic speed $V_a$, is distributed in the full diffusive halo, while it is restricted to the thin disc region in ours (for mathematical reasons, see \citealt{2020CoPhC.24706942M})\footnote{If CRs were to spend most of their time in the disc, the $V_a$ value in both models would be directly comparable. On the other hand, if CRs were to spend their time homogeneously at various halo heights, the re-acceleration $V_a$ in [De21] would be equivalent to a re-acceleration $V_a\times \sqrt{L/h}$ in ours \citep{2010A&A...516A..67M}. Reality lies in between these extreme cases.}. As a result, $V_a\sim29.9$~km~s$^{-1}$ (with $L_{\rm Webber}=1.1$~kpc) in [De21] corresponds to $V_a\lesssim99$~km~s$^{-1}$ in our model.
In Fig.~\ref{fig:impact_Va} we show calculations of $\mathrm{^{10}Be/^9Be}$ with and without re-acceleration, considering values up to 150~km/s. At 100~km/s (dotted red line), differences up to 20\% are observed, compared to the case with no re-acceleration. This is consistent with the results of \citet{2020JCAP...11..027Y}, where the authors find a $\sim 20\%$ larger halo size when considering re-acceleration (see their Table~1).
We do not wish to push further the comparison, but these numbers are in the range necessary to reconcile the difference in $L$ values between [De21] results (with re-acceleration) and those obtained from our analytical formula (without re-acceleration).

\end{document}